 \documentclass[aip, reprint, twocolumn, amsmath,amsfonts,amssymb]{revtex4-1}
\usepackage{graphicx}
\usepackage{amsmath,amsfonts,amssymb}
\usepackage{color}
\usepackage{bm}
\usepackage[mathscr]{eucal}
\usepackage{picture}
\usepackage{float}
\usepackage{subcaption}
\usepackage{braket}
\usepackage{graphicx}
\usepackage{amsmath,amsfonts,amssymb}
\usepackage{color}
\usepackage{bm}
\usepackage{threeparttable}
\usepackage[mathscr]{eucal}
\usepackage{picture}

\definecolor{grey}{rgb}{0.7,0.7,0.7}

\usepackage{fancyvrb}
\usepackage{ulem}

\usepackage[T1]{fontenc}

\begin{document}

\title{Many-Body-Expansion Based on Variational Quantum Eigensolver and Deflation for Dynamical Correlation}
\author{Enhua Xu}
\affiliation{Graduate School of System Informatics, Kobe University, 1-1 Rokkodai-cho, Nada-ku, Kobe, Hyogo 657-8501 Japan}

\author{Yuma Shimomoto}
\affiliation{Graduate School of System Informatics, Kobe University, 1-1 Rokkodai-cho, Nada-ku, Kobe, Hyogo 657-8501 Japan}

\author{Seiichiro L. Ten-no} 
\affiliation{Graduate School of System Informatics, Kobe University, 1-1 Rokkodai-cho, Nada-ku, Kobe, Hyogo 657-8501 Japan}

\author{Takashi Tsuchimochi} 
\email{tsuchimochi@gmail.com}
\affiliation{Graduate School of System Informatics, Kobe University, 1-1 Rokkodai-cho, Nada-ku, Kobe, Hyogo 657-8501 Japan}
\affiliation{Japan Science and Technology Agency (JST), Precursory Research for Embryonic Science and Technology (PRESTO), 4-1-8 Honcho Kawaguchi, Saitama 332-0012 Japan}

\begin{abstract}
In this study, we utilize the many-body expansion (MBE) framework to decompose electronic structures into fragments by incrementing the virtual orbitals. Our work aims to accurately solve the ground and excited state energies of each fragment using the variational quantum eigensolver and deflation algorithms. Although our approach is primarily based on unitary coupled cluster singles and doubles (UCCSD) and a generalization thereof, we also introduce modifications and approximations to conserve quantum resources in MBE by partially generalizing the UCCSD operator and neglecting the relaxation of the reference states. As a proof of concept, we investigate the potential energy surfaces for the bond-breaking processes of the ground state of two molecules ($\rm H_2O$ and $\rm N_2$) and calculate the ground and excited state energies of three molecules (LiH, CH$^+$, and $\rm H_2O$). The results demonstrate that our approach can, in principle, provide reliable descriptions in all tests, including strongly correlated systems, when appropriate approximations are chosen. Additionally, we perform model simulations to investigate the impact of shot noise on the total MBE energy and show that precise energy estimation is crucial for lower-order MBE fragments. 
\end{abstract}

\maketitle

\section{Introduction}
Finding an approximation to the full configuration interaction (FCI) solution for the many-electron electronic Schr\"odinger equation remains a challenging task in {\it ab initio} electronic structure theory, particularly for strongly correlated quantum systems. Several significant advancements have been made in the field of classical computation, such as the density matrix renormalization group method,\cite{DMRG_1, DMRG_2} stochastic approaches in configuration space,\cite{FCIQMC, i-FCIQMC, AFQMC} adaptive coupled cluster (CC) approach,\cite{FCCR, FCCR2} adaptive configuration interaction approaches\cite{SHCI, iCI, ASCI}, and the many-body-expansion FCI (MBE-FCI) approach.\cite{MBE_1,MBE_4} All of these methods can provide FCI-accuracy results for systems that cannot be processed using FCI. However, because the Hilbert space of quantum systems increases exponentially with the system size, the accuracy of these methods decreases significantly when the tested system becomes larger.

Quantum computation can naturally represent and manipulate quantum states; thus, it is expected to be an effective tool for simulating quantum mechanical systems.\cite{UseQM_Manin,UseQM_Fy,QM_simulator} Early implementations of quantum algorithms in computational chemistry\cite{PEA_2005} were deployed on quantum computers to evaluate molecular energies.\cite{VQE,QM_energy_2016,QM_energy_2017,QM_energy_2018} Unfortunately, quantum computers are extremely sensitive to errors caused by environmental disturbances, decoherence, and noise. Fault-tolerant quantum computing (FTQC) using quantum error-correcting codes was also investigated.\cite{FTQC_1, FTQC_2, FTQC_3, FTQC_4} However, currently, high-level encoding is not allowed due to restrictions on quantum resources such as qubit and magic state count.\cite{FTQC_2, FTQC_3} This limitation has been the driving force behind significant progress toward the development of algorithms that seek to reduce the quantum resources for solving quantum chemical problems. Some of these developments include quantum-classical hybrid algorithms for variational optimization\cite{VQE,classical_QM_1,classical_QM_2} and wavefunction ans\"atze that produce low-depth circuits for efficient quantum  simulation.\cite{Low_depth_PRX,Low_depth_PRL,Low_depth_JCTC,Low_depth_NC, Tsuchimochi20, Tsuchimochi22} 

Although quantum computing is expected to be suitable for capturing strong entanglements in a truncated orbital space, an accurate description of the electronic structure also requires the treatment of weak interactions owing to high-lying virtual orbitals. In quantum chemistry, these correlation effects are historically referred to as ``static correlation" and ``dynamical correlation,” respectively. The latter poses a significant challenge to quantum simulations because the number of qubits available for mapping the active orbital space is usually limited. Therefore, considerable effort has recently been devoted to developing preprocessing approaches that aim to compress a complicated electronic Hamiltonian into a compact active-space representation. Such approaches include double unitary coupled-cluster,\cite{Bauman19A, Bauman19B} transcorrelated,\cite{mcardle2020improving, sokolov2023orders, Tenno2023} and canonical transcorrelated methods.\cite{Yanai2012, Kumar2022}

Research has also focused on incorporating problem decomposition techniques developed for classical quantum chemistry applications \cite{PD_rev1, PD_rev2, PD_rev3, PD_rev4} into quantum algorithms to further improve the efficiency of simulations of near-term quantum and FTQC devices.\cite{Verma21, Fujii2022, Yoshikawa2022} Problem decomposition techniques offer the advantage of decomposing the full electronic structure problem of a molecule into smaller sub-problems that can be solved with fewer qubits and qubit gates, and the aforementioned pre-processed Hamiltonians can be easily combined. Such approaches have a long history, originating from the application of the Bethe-Goldstone equation \cite{BG_equation} to quantum chemistry in the 1960s.\cite{BG_equation_N1, BG_equation_N2, BG_equation_N3} It has been shown that, by decomposing the correlation energy of the molecular system into $n$-body sub-systems (increments) through the expansion of core \cite{iFCI_1, iFCI_2, iFCI_3} or secondary \cite{MBE_1, MBE_2, MBE_3} orbitals, correlation energies close to the FCI accuracy can be recovered at low values of $n$ in an embarrassingly parallel computation. The latter framework, denoted many-body-expansion FCI (MBE-FCI), was first proposed by Eriksen and Gauss in 2017.\cite{MBE_1}

In this study, we explore a strategy called MBE–VQE that combines the MBE framework and VQE to capture the residual correlation effect missed by the active space approach. A similar approach was used by Verma et al.\cite{Verma21} where the reference was Hartree-Fock (HF), and the many-body correlation was expanded by VQE using a unitary coupled-cluster with singles and doubles (UCCSD) ansatz.\cite{Trotter_3, UCCSD_VQE, DC_qUCC_VQE} However, we assume the presence of static correlation, for which HF is inadequate, and therefore use VQE within a small active space as the reference. Based on the VQE circuit, we increased the orbital space to estimate the dynamic correlation post-VQE. While it is possible to employ the same ansatz as the reference VQE to correlate with the fragment secondary orbitals, our scheme exploits the educated guess that the most essential description of dynamic correlation requires single and double particle-hole excitations from a reference wave function. Furthermore, we extend the scheme to excited states by adopting Variational Quantum Deflation (VQD),\cite{VQD} and demonstrate the feasibility of our approach by computing the vertical excitation energies of three small molecules (LiH, CH$^+$ and H$_2$O) as showcase examples. 

The remainder of this paper is organized as follows. Section II introduces the MBE-VQE (MBE-VQD) framework with the employment of the unitary coupled cluster (UCC) ansatz. In Section III, the computational details and results obtained for the tested systems are discussed. Section IV provides a summary of this study and the outlook for future development prospects.

\section{Theory}
\subsection{MBE Framework}
   
\begin{figure}[t!]
\begin{center}
    \includegraphics[width=0.5\textwidth]{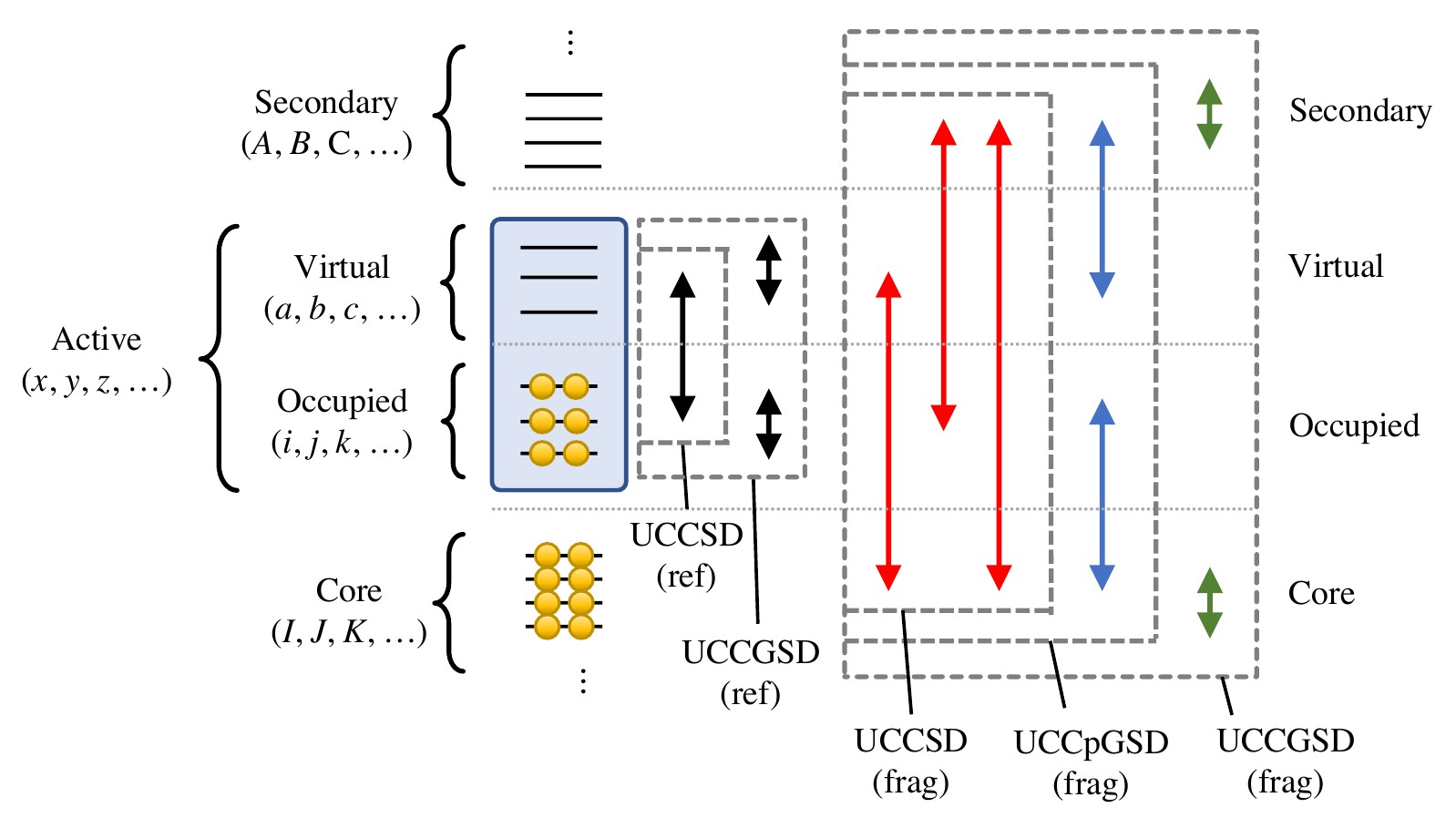}
\end{center}
\caption{Schematic illustration of the core, active, and secondary orbitals in the reference wavefunction. The arrows indicate the electron excitations considered in each method that correlate the corresponding spaces.}
\label{Orbitals}
\end{figure}

When describing strongly correlated systems, a linear combination of multiple determinants is typically employed as the reference wavefunction to provide a qualitatively correct description of the entire system. Based on the occupation of an orbital in the reference wavefunction (doubly occupied, partially occupied, or unoccupied), the complete set of spatial molecular orbitals (MOs) can be partitioned into three subsets: core, active, and secondary orbitals. In this study, we employ commonly used indices for the three subsets: $I,J,K,L$ for the core orbitals; $w, x, y, z$ for the active orbitals; and $A,B,C,D$ for the secondary orbitals, as illustrated in Figure \ref{Orbitals}. The active space can be further decomposed into active occupied ($i,j,k,l$) and active virtual ($a,b,c,d$) orbitals, given that the initial quantum state is typically an HF state.

In the MBE-FCI method, a complete set of MOs is divided into reference and expansion spaces. In general, the reference space consists of active orbitals, and each fragment includes all active orbitals and a growing number of MOs from the expansion space, which consists of core and secondary orbitals. The residual correlation energy ($E_{\rm corr}$), defined as the deviation of the reference energy ($E_{\rm ref}$) from the FCI ($E_{\rm FCI}$), can be expressed in terms of $n$body increments as follows:\cite{MBE_4, MBE_5, MBE_6, MBE_7} 
		\begin{equation}
        \begin{aligned}
            E_{\rm corr} &= E_{\rm FCI} - E_{\rm ref} \\
            &= \sum_{p}\Delta \epsilon_{p} + \sum_{p>q}\Delta \epsilon_{pq} + \sum_{p>q>r}\Delta \epsilon_{pqr} + \cdots \\
            &= E^{(1)} + E^{(2)} + E^{(3)}  + \cdots.
        \end{aligned}
        \label{eq:Ecorr}
		\end{equation}
Here, $p, q, r$ are the general indices representing the core or secondary orbitals. $\epsilon_{p} \equiv \Delta \epsilon_p$ is the correlation energy obtained by performing a calculation on fragment $p$, which is a composite space of orbital $p$ and active reference space. Similarly, $\epsilon_{pq}$ is the correlation energy from fragment $p, q$. The actual two-body increment of adding orbitals $p$ and $q$ into the reference space simultaneously (denoted as $\Delta \epsilon_{pq}$) is the increment obtained by subtracting $\epsilon_{p}$ and $\epsilon_{q}$ from $\epsilon_{pq}$:
		\begin{equation}
        \begin{aligned}
            \Delta \epsilon_{pq} = \epsilon_{pq} - (\Delta \epsilon_{p} + \Delta \epsilon_{q})
        \end{aligned}
        \label{eq:delta_epq}
		\end{equation}
Similarly, the three-body increment $\Delta \epsilon_{pqr}$ is computed as follows:
		\begin{equation}
        \begin{aligned}
            \Delta \epsilon_{pqr} = \epsilon_{pqr} - (\Delta \epsilon_{p} + \Delta \epsilon_{q} + \Delta \epsilon_{r}) - (\Delta \epsilon_{pq} + \Delta \epsilon_{pr} +\Delta \epsilon_{qr})
        \end{aligned}
		\end{equation}
Generally, if the active orbitals are well chosen, most of the static correlation is described in the reference wavefunction, and the remaining dynamic correlation is obtained by calculating the increments of the generated fragments. Consequently, the MBE expansion in Eqs. (\ref{eq:Ecorr}) converges rapidly. In contrast, an insufficient reference wavefunction may lead to oscillations in the subsequent MBE energies, as discussed in Ref. [\onlinecite{MBE_3}].

From the definition of the increments above, it is clear that the calculation of higher-order approximations to $E_{\rm FCI}$ presupposes knowledge of the components of all contributions at lower orders, in the sense that lower-order increments enter expressions for higher-order increments. Thus, if an error is introduced into the reference energy or a lower-order increment, it is passed on to the corresponding higher-order increments and accumulates in the total energy. 

To illustrate the severity of this error accumulation, let us assume that there are $M_s$ secondary orbitals and no core orbitals and that the correlation energies of all the fragments are exactly solved, except for $\epsilon_A$, which is slightly larger than it should be (suppose this error is $\delta_A$). The accuracy of $E^{(1)}$ can be assured as it is lower than its exact value by only $\delta_A$. However, according to Eq. (\ref{eq:delta_epq}), all the two-body increments related to orbital $A$ $(\Delta \epsilon_{AB}$, $\Delta \epsilon_{AC}, ....)$ will be $\delta_A$ smaller, resulting in $E^{(2)}$ being higher than the exact value by $\delta_A(M_s-1)$. By analogy, $E^{(3)}$ will be decreased by a much larger magnitude. 

This may not be a serious problem in MBE-FCI as all fragment energies may be precisely calculated (assuming that there is no approximation in the FCI solver and round-off errors can be ignored). However, considerable quantum noise in quantum computing, including shot noise, is inevitable at this stage, and even in the foreseeable future. To control its effect on MBE, we consider the following two measures.

The first measure involves freezing the core orbitals and considering only the correlation between the active and secondary orbitals. This is in agreement with earlier MBE-FCI work\cite{MBE_1, MBE_2, MBE_3}, which was proven to be able to converge at lower orders. Because the error accumulation grows very rapidly as the order increases, such a treatment can effectively control its effect on the total energy.

The second measure  entails reducing the shot noise by increasing the number of measurements on the reference and lower-order fragments. In variational quantum computing, the energy expectation value is obtained from the average value of many measurements. Therefore, by performing more measurements, the shot noise can be reduced, but this requires more time. As mentioned above, the noise in the reference or lower-order fragments accumulates more severely than that in the higher-order fragments. Thus, it is cost-effective to perform more measurements on the reference and lower-order fragments, which are also much smaller in number than the high-order fragments. At the end of Section III, we present  a model simulation to demonstrate that this treatment is highly efficient.

\subsection{UCC Ansatz}
The UCC ansatz takes the exponential form of the anti-hermitized operators $\hat S$,
\begin{align}
    &\hat S = \hat S_1 + \hat S_2 + \cdots\\
    &\hat S_1 = \sum_{p_1 q_1} t^{p_1}_{q_1} \left(a^\dag_{p_1}a_{q_1} -a^\dag_{q_1}a_{p_1}\right)\\
    &\hat S_2 = \sum_{p_1 p_2q_1 q_2} t^{p_1p_2}_{q_1q_2} \left(a^\dag_{p_1}a^\dag_{p_2}a_{q_2}a_{q_1} -a^\dag_{q_1}a^\dag_{q_2}a_{p_2}a_{p_1}\right)\\
    & \cdots\nonumber
\end{align}
where $t_{q_1}^{p_1}, t_{q_1 q_2}^{p_1p_2}, \cdots$ are real parameters and the definition of the orbital spaces for $p_1,\cdots$ depends on the method. 
The most common approach is UCCSD, where $\hat S$ contains only single and double substitutions, whose excitations and de-excitations occur only between the occupied and virtual orbitals, like the traditional single-reference CC method (see Fig.\ref{Orbitals}).

Another well-known approach is the UCC with generalized singles and doubles (UCCGSD) ansatz. Unlike UCCSD, the orbital indices in UCCGSD are not limited to occupied or virtual orbitals but are generalized to include all orbitals. Therefore, UCCGSD has more parameters and can provide more accurate results, although it also incurs a higher computational cost. Interestingly, the number of generalized singles and doubles parameters is equal to the number of parameters in the Hamiltonian. As a result, the expectation once existed that the Schr\"odinger equation could be solved exactly by the coupled cluster method with generalized singles and doubles (CCGSD),\cite{CCGSD_Nooijen, CCGSD_Nakatsuji, CCGSD_Troy, CCGSD_Piecuch} although this expectation was later proved to be incorrect.\cite{CCGSD_Davidson, CCGSD_Nonen, CCGSD_Mukherjee} Nevertheless, CCGSD and UCCGSD do deliver much higher performances than CCSD and UCCSD, particularly for describing strongly correlated systems.\cite{CCGSD_Troy, CCGSD_Piecuch, UCCGSD}

\begin{figure}[htbp]
\begin{center}
    \includegraphics[width=0.48\textwidth]{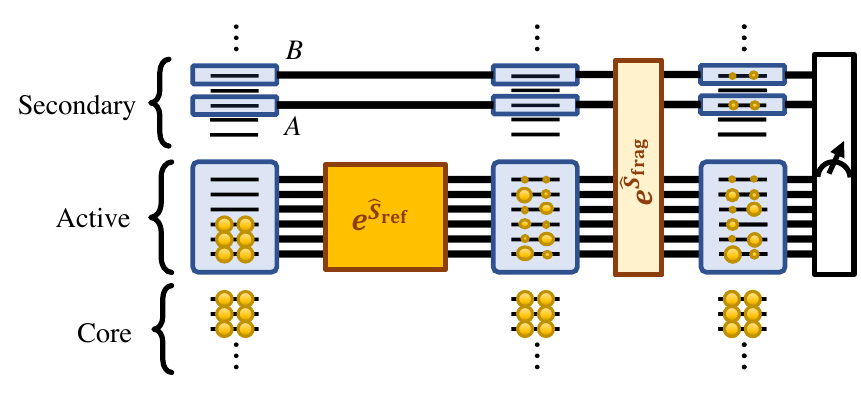}
\end{center}
\caption{Schematic quantum circuit diagram of the VQE algorithm for $\ket {\Psi_{\rm frag}}$ to compute $\epsilon_{AB}$. The initial state is set to a product state representing a single determinant, denoted as $|\Phi\rangle$, whose active space is correlated by the parameterized quantum circuit of $e^{\hat S_{\rm ref}}$. This is then further correlated with orbitals $A$ and $B$ by $e^{\hat S_{\rm frag}}$, followed by the measurements to estimate the energy expectation value.}
\label{circuit}
\end{figure}

In our MBE-UCC method, the generation of the reference wavefunction ($\ket {\Psi_{\rm ref}}$) is formulated as:
		\begin{equation}
        \begin{aligned}
            \ket {\Psi_{\rm ref}} = e^{\hat S_{\rm ref}} \ket {\Phi}. 
        \end{aligned}\label{eq:ref}
		\end{equation}
Here, $\hat S_{\rm ref}$ is a specified UCC operator and $\ket {\Phi}$ is set as one determinant in this study.  The parameters in $\hat S_{\rm ref}$ are optimized to obtain the optimal solution of $\ket {\Psi_{\rm ref}}$. Then, the fragment wavefunction ($\ket {\Psi_{\rm frag}}$) is then built upon the optimized $\ket {\Psi_{\rm ref}}$ as follows:
  		\begin{equation}
        \begin{aligned}
            \ket {\Psi_{\rm frag}} = e^{\hat S_{\rm frag}} \ket {\Psi_{\rm ref}} = e^{\hat S_{\rm frag}} e^{\hat S_{\rm ref} } \ket {\Phi}
        \end{aligned}
		\end{equation}
where $\hat S_{\rm frag}$ is also a UCC operator that can be either the same ansatz as or a different ansatz from $\hat S_{\rm ref}$. Since $\ket {\Psi_{\rm ref}}$ is considered multi-determinant, this is the unitary version of the so-called ``internally-contracted'' multi-reference coupled cluster.\cite{IC_MRCC_Kohn, IC_MRCC_Mukherjee, IC_MRCC_Yanai, IC_MRCC_Evangelista} The quantum circuit for the generation of $\ket {\Psi_{\rm frag}}$ is illustrated in Figure \ref{circuit}. In our MBE-UCC work, 
the number of secondary orbitals used in the fragment (denoted as $m$) is at most three. Thus, the number of active orbitals (denoted as $N$) can be easily much larger than $m$ and the corresponding quantum resources, such as qubits and qubit gates related to $e^{\hat S_{\rm ref}}$ will be much greater than those related to $e^{\hat S_{\rm frag}}$.

It should be noted that the parameters {\bf t} in $e^{\hat S_{\rm ref}}$ can be fixed or re-optimized in the generation of $\ket {\Psi_{\rm frag}}$, which are denoted as reference-unrelaxed or reference-relaxed schemes, respectively. The former can be considered an approximation to the latter one.

If $\hat S_{\rm ref}$ and $\hat S_{\rm frag}$ are both UCCSD (UCCGSD) operators, the resulting approach is denoted MBE-UCCSD (MBE-UCCGSD). Upon convergence, these methods are expected to provide the energy of the corresponding method for a large basis set.

Although the framework of the original MBE is well established and is expected to be transferable to VQE, we aim to reduce the cost of both quantum resources and classical optimization in VQE by introducing the following approximations to make the approach even more feasible: the use of a single trotter step, partial generalization of the UCCSD operator, and fixing the reference $|\Psi_{\rm ref}\rangle$ in the fragment computation.

\subsubsection{\rm Single Trotter Step}
The first approximation concerns the implementation of the exponential ansatz, $e^{\hat S_{\rm frag}}$ and $e^{\hat S_{\rm frag}}$ as a quantum circuit. In order to make the UCC operator programmable on a quantum device,  Trotterization is required for noncommutative exponents.\cite{Trotter_1, Trotter_2, Trotter_3} Applying UCCSD or UCCGSD to strongly correlated systems often triggers large $t$ amplitudes to account for higher excitations, which, in turn, necessitates a large Trotter number $\mu$, thereby requiring a high-depth quantum circuit. Fortunately, many authors have found that the Trotter approximation with $\mu = 1$ can be used, because the variational optimization of the VQE energy can absorb the Trotter error even at $\mu = 1$.\cite{Trotter_1, Trotter_4, Trotter_5} This is the so-called single Trotter step,\cite{Trotter_1} and all the calculations in this work are implemented with this Trotter approximation.

\begin{figure}[t!]
\begin{center}
    \includegraphics[width=0.45\textwidth]{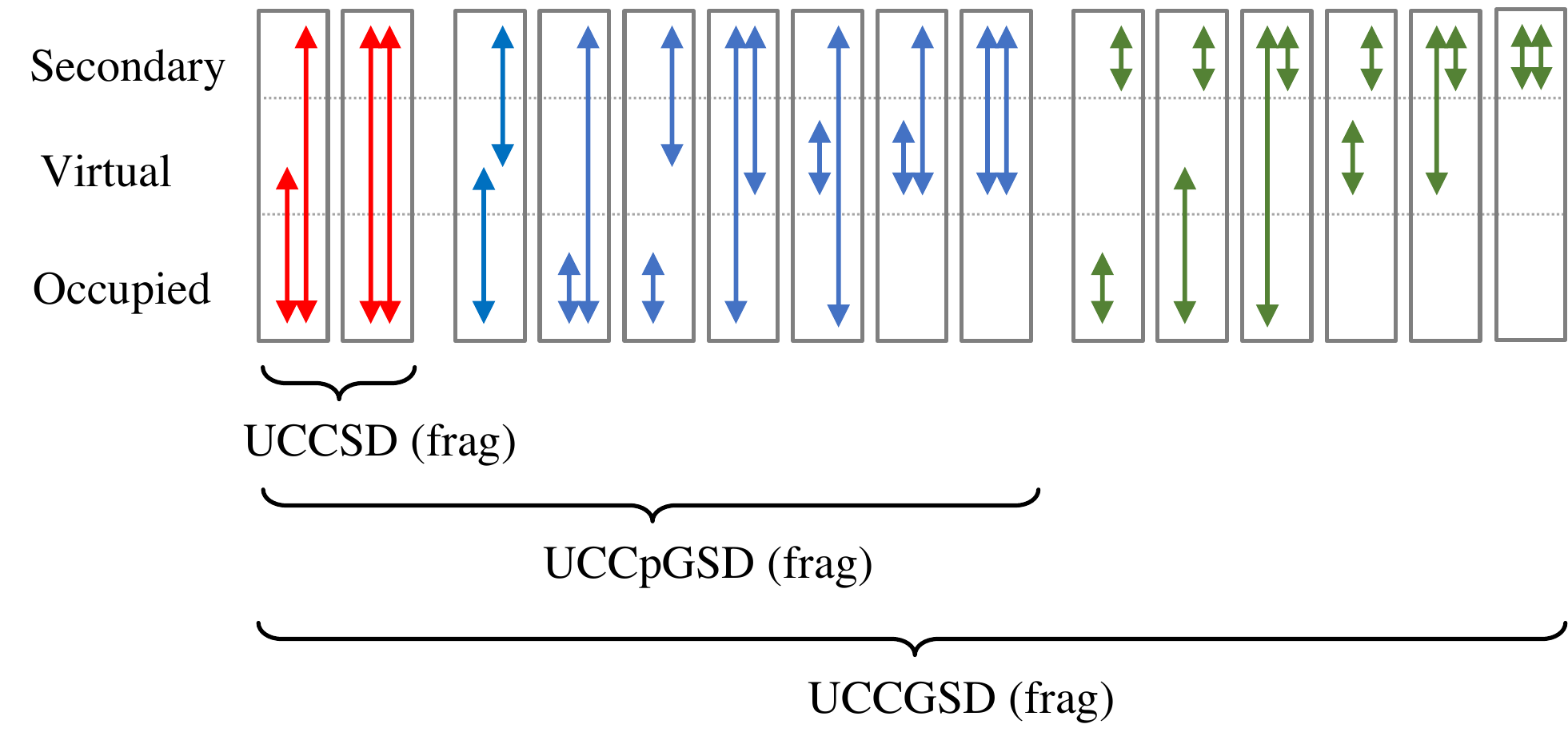}
\end{center}
\caption{Types of double substitutions that are considered in $\hat S_{\rm frag}$ for each method. The core space is not presented for simplicity.}
\label{Orbitals2}
\end{figure}

\subsubsection{Partially Generalized UCCSD Operator} \label{sec:UCCpGSD}
The second approximation concerns the simplification of $e^{\hat S_{\rm frag}}$ in the MBE-UCCGSD. As mentioned previously, fragment computation is expected to capture the dynamical correlation missing from $\ket {\Psi_{\rm ref}}$. Therefore, we consider it possible to use a modified UCCGSD operator instead of the costly UCCGSD operator for $\hat S_{\rm frag}$ in MBE-UCCGSD, in which the occupied orbital indices run over the core and active orbitals, whereas the virtual orbital indices run over the active and secondary orbitals, except for excitations in the active space, as shown in Figure \ref{Orbitals}. For clarity, we summarize the double substitutions considered for each method in Figure \ref{Orbitals2} (the core space is excluded for simplicity). This modification can be considered as a kind of ``partially generalized" UCCSD (UCCpGSD),\cite{MRUCCSD} an approach usually exercised in internally-contracted multi-reference methods, and the resulting MBE approach is denoted as MBE-UCCGSD/UCCpGSD. 

We also considered employing the UCCpGSD operator $\hat S_{\rm frag}$ in MBE-UCCSD. However, this MBE-UCCSD/UCCpGSD results in severe inconsistency in the treatment of the correlation effect between $\ket {\Psi_{\rm ref}}$ and $\ket {\Psi_{\rm frag}}$ because the latter can refine the description of the active space in the presence of active-to-active-type excitations (blue arrows in Figure \ref{Orbitals2}). Specifically, compared with MBE-FCI, MBE-UCCSD/UCCpGSD provides fragment total energies that are relatively more accurate than the reference UCCSD energy $E_{\rm ref}$. This leads to less accurate and overestimated fragment {\it correlation} energies $\epsilon_p$, $\epsilon_{pq}$, etc.

In contrast, the inconsistency resulting from the partial generalization in MBE-UCCGSD/UCCpGSD is expected to be much smaller because the correlation effect missing from the method is typically negligible; core-to-core and secondary-to-secondary excitations in MBE-UCCGSD (green arrows in Figure \ref{Orbitals2}) make negligible contributions to the correlation energy if the active space is well chosen. We return to this point in Section \ref{sec:comparison}.

\subsubsection{\rm Reference-unrelaxed approximation}
\begin{figure}[t!]
\begin{center}
    \includegraphics[width=0.47\textwidth]{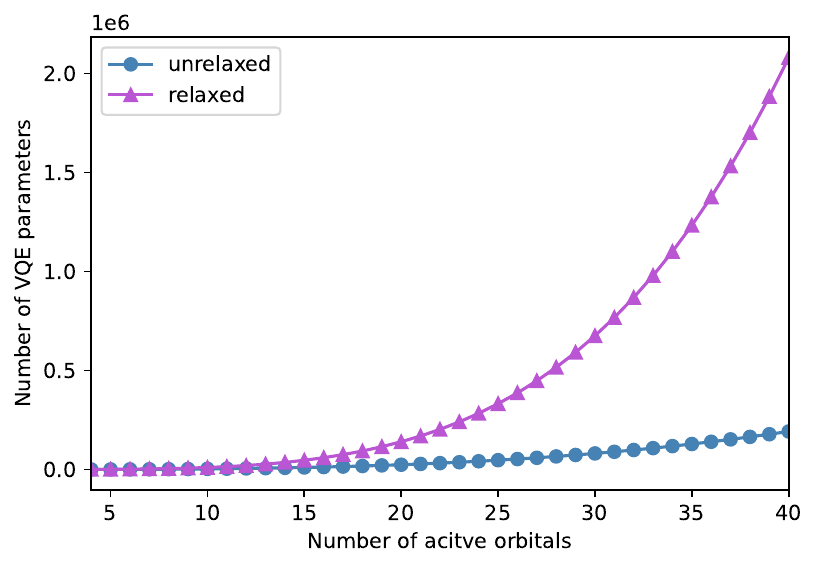}
\end{center}
\caption{Number of VQE parameters under the reference-relaxed and reference-unrelaxed approaches in the fragment that includes only one secondary orbital with respect to a different number of active orbitals.}
\label{relax_unrelax}
\end{figure}

The third is the reference-unrelaxed approximation mentioned above, which can lower the cost of optimizing $e^{\hat S_{\rm ref}}$ when computing the fragment. When the number of active orbitals $N$ is much greater than the number of additional fragment orbitals $m$, $N\gg m$, the reference-unrelaxed scheme can significantly reduce the number of UCC parameters that would need to be optimized in MBE-UCCGSD/UCCpGSD from $O(N^2(N+m)^2)$ to $O(2mN^3)$. Figure \ref{relax_unrelax} shows the number of VQE parameters under the reference-relaxed and reference-unrelaxed approaches with respect to different numbers of active orbitals $N$ for fragment calculation with one secondary orbital ($m=1$). As is evident from the figure, the larger the active space, the higher the proportion of VQE parameters that can be saved. For example, in a situation with more than 37 active orbitals, the reference-unrelaxed approximation can save more than 90$\%$ of the parameters. Such substantial savings in the optimization cost would be highly beneficial to the stability of quantum computing. It should be mentioned that the number of parameters in the reference-unrelaxed MBE-UCCGSD scales as $O(4mN^3)$, which indicates that under this approximation, the UCCpGSD operator would lower the optimization cost to approximately half of its original value.

We realized that the reference-unrelaxed approximation can occasionally introduce some error in the fragment correlation energy and also found that this error can be significantly suppressed by employing optimized orbitals. This is discussed in detail in Section III.

\begin{table*}[t!]
\tabcolsep=0.5em
	\caption{\label{tab:CH+_TEST}
 		Ground state energies of CH$^+$ with the aug-cc-pVDZ basis set provided by different approaches. The MBE-FCI energies are in a.u., whereas the deviations of other approaches from MBE-FCI are in mH (milliHartree).}
	\begin{tabular}{cccccccc}
		\hline\hline
		     & FCI & UCCSD & UCCSD/UCCpGSD   & UCCGSD & UCCGSD/UCCpGSD \\
		\hline
  		Ref    & -37.925641 & 0.008 & 0.008    &  0.000 & 0.000  \\
		MBE(1) & -37.973872 & 0.360 & -0.190   & 0.000 & 0.000  \\
  		MBE(2) & -38.005191 & 1.053 & 2.276    & 0.000 & -0.001 \\
		MBE(3) & -38.005066 & 1.667 & -17.429  & 0.000 & 0.004  \\
  		\hline\hline
		     
	\end{tabular}
\end{table*}
\subsection{Excited states}
In the framework of variational quantum algorithms, considerable efforts have been made to obtain excited states.\cite{McClean17, Colless18,Higgott19, Nakanishi19, Ollitrault20, Zhang21A,  Tkachenko22, Heya23, Tsuchimochi23A, Tsuchimochi23B, Yoshikura23} Among them, the most prominent approach is perhaps variational quantum deflation (VQD),\cite{Higgott19} which introduces the following modification to the Hamiltonian to determine the $\kappa$th excited state:
		\begin{equation}
        \begin{aligned}
            \hat H^{(\kappa)} = \hat H + \sum_{\lambda=0}^{\kappa-1} \alpha \ket{\Psi_\lambda} \bra{\Psi_\lambda} 
        \end{aligned}
		\end{equation}
This modified Hamiltonian ensures approximate orthogonality between $|\Psi_\lambda\rangle$  $(\lambda=0,1,2,\cdots, \kappa)$, which are the ground state, and the first, second, $\cdots$, and $\kappa$th excited states,  obtained by the corresponding VQE calculations. 

MBE-VQE can be easily generalized to MBE-VQD for excited-state calculations by replacing the Hamiltonian with $\hat H^{(\kappa)}$. However, one caveat is that the HF orbitals typically employed for quantum simulations, such as VQE (and VQD), are optimized with respect to the ground state but not for excited states. This leads to an imbalance between the ground and excited states, resulting in a qualitatively incorrect description of the excitation energies at low MBE orders.

To address this issue, we introduced orbital optimization in the reference VQE and VQD states. The orbital-optimized (oo) VQE is similar to the complete active space self-consistent field (CASSCF) and can be classically performed using first- and second-order reduced density matrices.\cite{Sokolov20, Mizukami20} To generalize the orbital optimization to VQD, we simply take the state-average approach where the sum of the VQE and VQD energies is minimized by appropriate orbital rotation. As shown below, with optimized orbitals, the ground- and excited-state energies converge faster than those with HF orbitals with respect to the order of MBE.

\section{Results}
\subsection{Computational Details}
Both MBE-FCI and MBE-UCC were implemented in our Python-based emulator package {\sc Quket},\cite{quket} which complies with other useful libraries such as {\sc Openfermion},\cite{OPENFERMION} {\sc Pyscf},\cite{PYSCF} and {\sc Qulacs}.\cite{qulacs} The Jordon-Wigner transformation\cite{JW_trans} was employed to map the fermion operators to the qubit representation, and the tapering-off technique for reducing the number of qubits was used.\cite{bravyi17, Setia20} We did not consider quantum noise in our calculations. In some calculations, the orbitals (including the core orbitals) were optimized. For cases where FCI is unavailable, the multi-reference configuration interaction with Davidson correction (MRCI+Q) energies\cite{MRCI_1, MRCI_2, Pople} based on either the HF or CASSCF orbitals were used as the benchmark. In the following applications, we employ MBE-VQE to describe the potential energy surfaces (PESs) for the bond dissociation processes of the ground state of two molecules (H$_2$O and N$_2$), and employ both MBE-VQE and MBE-VQD to calculate the ground- and excited-state energies of three molecules (LiH, CH$^+$ and H$_2$O). All the data shown below were computed using the reference-unrelaxed approximation, unless otherwise noted. The results of the reference-relaxed scheme are listed in the Supplementary Information (SI).

\begin{figure*}[t!]
\includegraphics[width=0.99\textwidth]{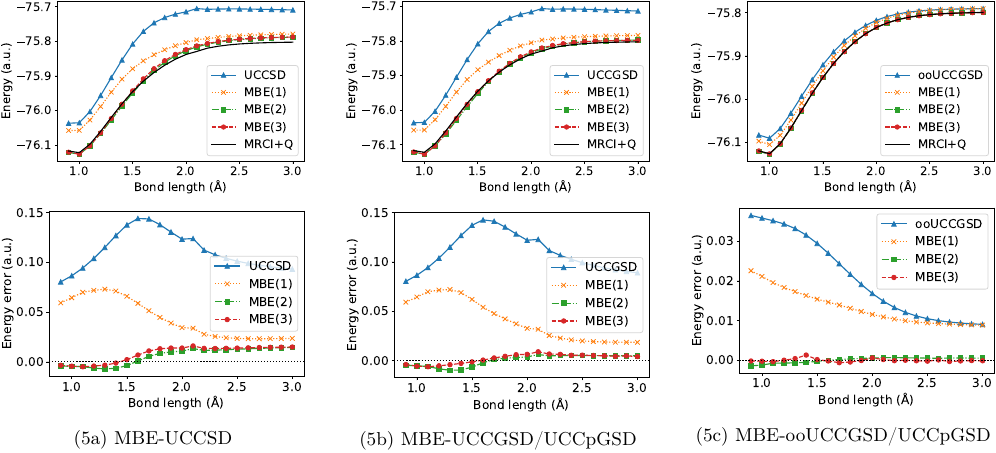}
     \caption{Absolute and relative (with respect to MRCI+Q) energies of different approaches in the bond dissociation process of the ground state of H$_2$O with the aug-cc-pVDZ basis set.  }
     \label{fig:H2O_PES}
\end{figure*}

\subsection{Comparison of MBE methods}\label{sec:comparison}
We first investigate the performance of MBE using different ans\"atze using the ground state of the CH$^+$ molecule as an example. The bond length is 1.131 {\AA} and we use the HF orbitals with the aug-cc-pVDZ basis set. The C$1s$ core orbital is frozen and the next five orbitals are selected as the active orbitals. Table \ref{tab:CH+_TEST} presents the ground state energies calculated by MBE($n$)-FCI and the energy differences of the four proposed "reference-relaxed" MBE($n$)-UCC approaches, where $n$ indicates the order of expansion.

As shown in the table, MBE(3)-FCI almost converges, given that the FCI energy is $-38.004247$ Hartree. The MBE(1)-UCCSD/UCCpGSD energy is slightly lower than that of MBE(1)-FCI because of the slightly larger fragment correlation energy $\{\Delta \epsilon_A\}$. As indicated in Section \ref{sec:UCCpGSD}, the second- and third-order MBE-UCCSD/UCCpGSD energies increasingly deviate from the MBE-FCI results in opposite directions, implying the divergence of the higher-order MBE energies. In contrast, the energies of MBE-UCCSD did not oscillate because of the consistent treatment of the correlation effects, with a reasonably accurate energy of $-38.003399$ Hartree at third order. 

Both MBE-UCCGSD and MBE-UCCGSD/UCCpGSD performed quite well for this system; the accuracy of the former is essentially equivalent to that of MBE-FCI, and the largest deviation of the latter from MBE-FCI is a mere 0.004 mH at the third order. Notably, the energy oscillation in MBE-UCCGSD/UCCpGSD is drastically suppressed compared to that of MBE-UCCSD/UCCpGSD, and its performance is almost the same as that of MBE-UCCGSD. This clearly demonstrates that the missing correlation effect in UCCpGSD, that is, secondary-to-secondary excitations, has almost no contribution and can be practically neglected. Therefore, in the remainder of this paper, we report and compare only the results of MBE-UCCSD and MBE-UCCGSD/UCCpGSD.

\subsection{Potential Energy Surface}
\subsubsection{\rm H$_2$O}
Next, we examine the bond-breaking process of H$_2$O using the aug-cc-pVDZ basis set. The H-O-H angle was fixed at $104.33^\circ$ and the two O-H bonds were varied from 0.9 \r{A} to 3.0 \r{A}. We froze the $1a_1, 2a_1,$ and $1b_1$ orbitals and chose the $3a_1, 1b_2, 4a_1,$ and $2b_2$ orbitals as the active space. The absolute energies and their deviations from MRCI+Q of the different methods are depicted in Figure \ref{fig:H2O_PES}. As illustrated in Figures \ref{fig:H2O_PES}a and \ref{fig:H2O_PES}b, MBE-UCCSD and MBE-UCCGSD/UCCpGSD exhibit similar behavior at the reference and first order. However, for the second and third orders, MBE-UCCGSD/UCCpGSD is notably closer to MRCI+Q than MBE-UCCSD is, particularly in the bond-breaking region. This highlights the superior ability of MBE-UCCGSD/UCCpGSD compared with MBE-UCCSD in describing strongly correlated systems. The accuracy and convergence of MBE-UCCGSD/UCCpGSD can be further improved by optimizing the orbitals during the reference UCCGSD calculation, which is denoted as ooUCCGSD. Accordingly, the benchmark MRCI+Q is based on CASSCF orbitals. The improvement in MBE-ooUCCGSD/UCCpGSD over MBE-UCCGSD/UCCpGSD was significant, as shown in Figure 4c. We present the maximum absolute error (MAE) and non-parallelity error (denoted as NPE, which refers to the difference between the maximum and minimum deviations) with respect to MRCI+Q for the MBE(2) and MBE(3) data in Table \ref{tab:H2O_PES}. Surprisingly, even MBE(2)-ooUCCGSD/UCCpGSD exhibits much smaller MAE and NPE values than MBE(3)-UCCSD and MBE(3)-UCCGSD/UCCpGSD. The overwhelming advantage of MBE-ooUCCGSD/UCCpGSD illustrates the effectiveness of orbital optimization.

\begin{table}[b!]
	\caption{\label{tab:H2O_PES}
 		MAE and NPE with respect to MRCI+Q in the PES for the bond dissociation process of the ground state of H$_2$O. The aug-cc-pVDZ basis set is employed and all the data are in mH (milliHartree).}
	\begin{tabular}{ccc}
		\hline
		     & MAE & NPE \\
		\hline\hline
  		MBE(2)-UCCSD & 14.5 & 21.5 \\
		MBE(3)-UCCSD & 16.0 & 20.9 \\
  		MBE(2)-UCCGSD/UCCpGSD & 9.7 & 15.9  \\
		MBE(3)-UCCGSD/UCCpGSD & 9.0 & 15.1  \\
     	MBE(2)-ooUCCGSD/UCCpGSD & 1.4 & 2.1 \\
   		MBE(3)-ooUCCGSD/UCCpGSD & 1.3 & 1.9 \\
        \hline
	\end{tabular}
\end{table}

\begin{figure*}[t!]
\includegraphics[width=0.99\textwidth]{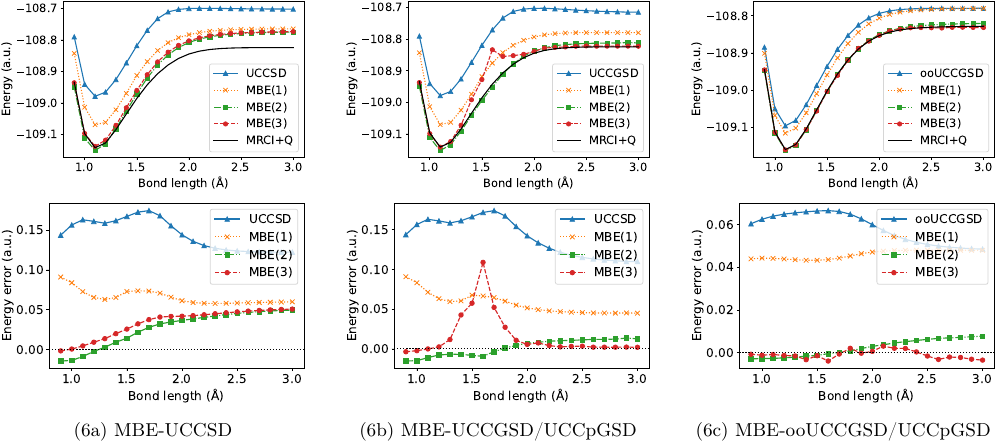}
     \caption{Absolute and relative (with respect to MRCI+Q) energies of different approaches in the bond dissociation of the ground state of N$_2$ with the aug-cc-pVDZ basis set.}\label{fig:N2_PES}
\end{figure*}

\subsubsection{\rm N$_2$}
The PES of the triple bond breaking process of the ground state of N$_2$ is more challenging to describe accurately than that of H$_2$O. The aug-cc-pVDZ basis set was employed and the N-N bond was stretched from 0.9 \r{A} to 3.0 \r{A}. The $1\sigma_g, 2\sigma_g, 1\sigma_{u}$ and $2\sigma_{u}$ orbitals were frozen and the $1\pi_u, 2\pi_{u}, 3\sigma_{g}, 1\pi_{g}, 2\pi_{g},$ and $3\sigma_{u}$ orbitals were chosen to be active. According to Figures \ref{fig:N2_PES}a and \ref{fig:N2_PES}b, the performances of MBE-UCCSD and MBE-UCCGSD/UCCpGSD are similar to those in Figures \ref{fig:H2O_PES}a and \ref{fig:H2O_PES}b, except that the MBE(3)-UCCGSD/UCCpGSD energies around 1.6 \(\text{\AA}\) are quite different from the benchmark MRCI+Q energies. As listed in Table \ref{tab:N2_PES}, the MAE and NPE values of MBE(3)-UCCGSD/UCCpGSD are as high as 108.9 and 112.5 mH, respectively. However, as shown in Figure S2b and Table SII in SI, the PES of the reference-relaxed MBE(3)-UCCGSD/UCCpGSD is very close to the benchmark results, where the MAE and NPE values are only 3.3 and 6.5 mH, respectively. As they are both performed on the same HF orbitals, such abnormally different performances can only be attributed to the reference-unrelaxed approximation. 

On the other hand, using the optimized orbitals, the PES of MBE(3)-ooUCCGSD/UCCpGSD in Figure \ref{fig:N2_PES}c becomes very close to the benchmark, and its MAE and NPE values are only 3.9 and 7.0 mH, respectively. This performance demonstrates that the error caused by the reference-unrelaxed approximation is mainly attributed to the inadequate description of the HF orbitals, especially when the system is strongly correlated, and can be significantly suppressed by employing the optimized orbitals.

\begin{table}[t!]
	\caption{\label{tab:N2_PES}
 		MAE and NPE with respect to MRCI+Q in the PES of the bond dissociation of the ground state of N$_2$. The aug-cc-pVDZ basis set is employed and all the data are in mH}
	\begin{tabular}{ccc}
		\hline
		     & MAE & NPE \\
		\hline\hline
  		MBE(2)-UCCSD & 49.5 & 64.2 \\
		MBE(3)-UCCSD & 50.7 & 52.6 \\
  		MBE(2)-UCCGSD/UCCpGSD & 15.3 & 28.9  \\
		MBE(3)-UCCGSD/UCCpGSD & 108.9 & 112.5  \\
     	MBE(2)-ooUCCGSD/UCCpGSD & 7.8 & 10.5 \\
   		MBE(3)-ooUCCGSD/UCCpGSD & 3.9 & 7.0 \\
        \hline
	\end{tabular}
\end{table}

\begin{figure*}[t!]
\begin{center}
    \includegraphics[width=0.99\textwidth]{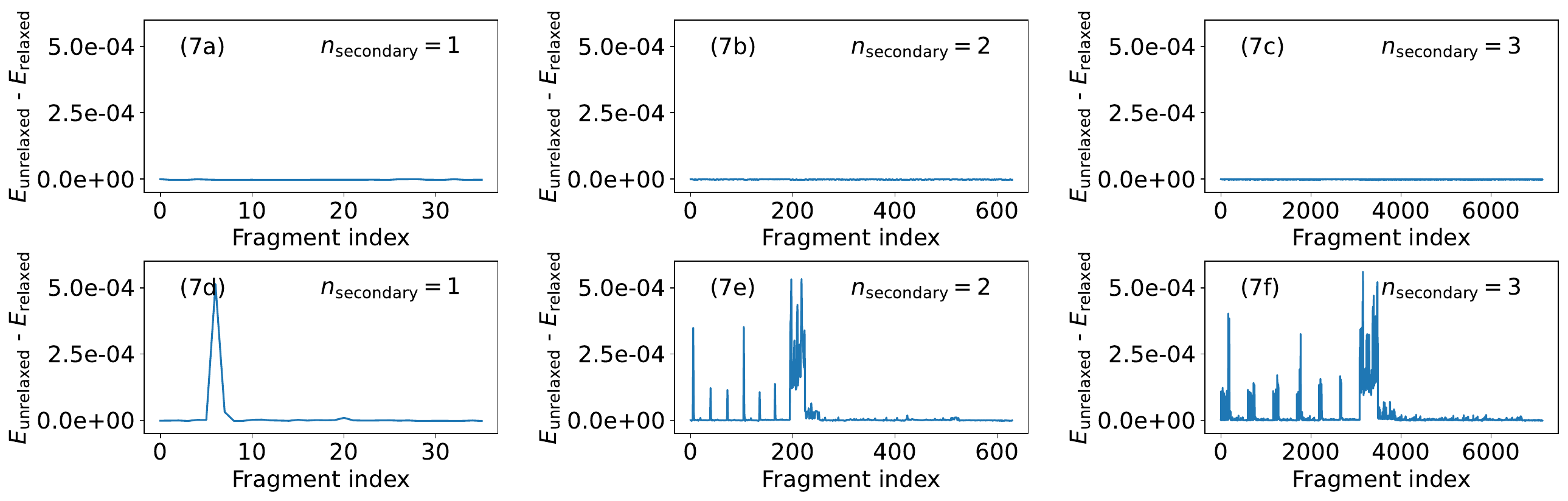}
\end{center}
\caption{Energy deviations of each fragment in MBE-UCCGSD/UCCpGSD between reference-unrelaxed and reference-relaxed of N$_2$ at 1.0 \(\text{\AA}\) and 1.6 \(\text{\AA}\) based on HF orbitals.}
\label{HF_deviations}
\end{figure*}

\begin{figure*}[t!]
\begin{center}
    \includegraphics[width=0.99\textwidth]{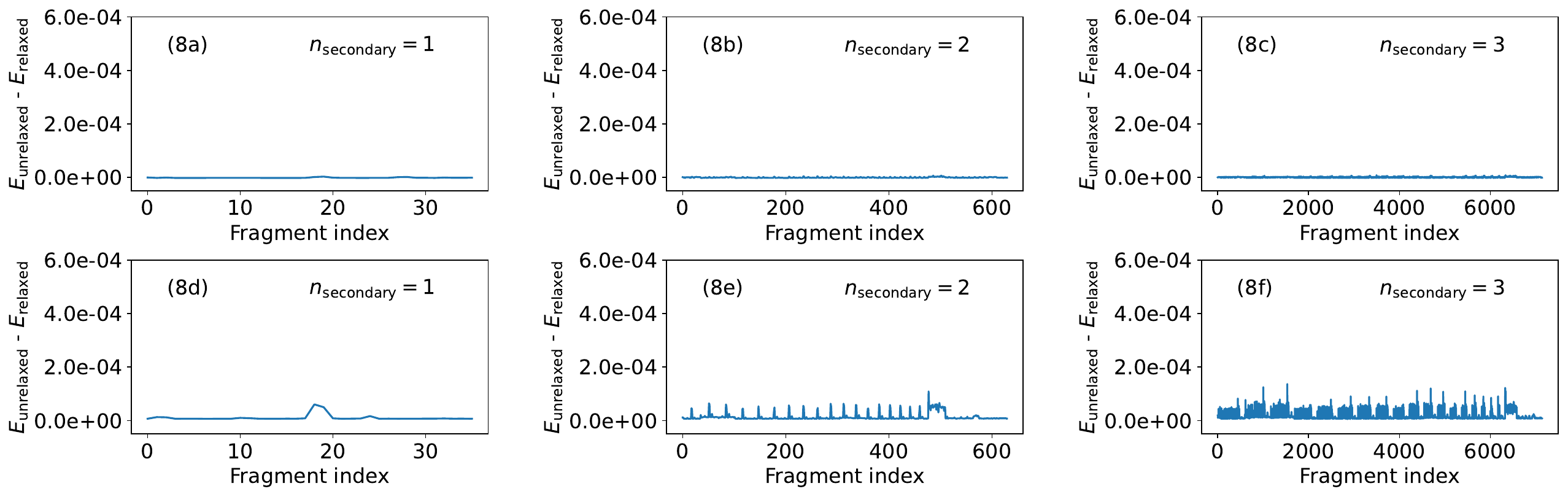}
\end{center}
\caption{Energy deviations of each fragment in MBE-UCCGSD/UCCpGSD between the reference-unrelaxed and reference-relaxed approaches of N$_2$ at 1.0 \(\text{\AA}\) and 1.6 \(\text{\AA}\) with optimized orbitals.}
\label{OPT_deviations}
\end{figure*}

To enable a clear visualization of the errors from the reference-unrelaxed approximation, we plot the energy deviations of each fragment between the reference-unrelaxed and reference-relaxed approaches in MBE-UCCGSD/UCCpGSD in Figure \ref{HF_deviations} (with HF orbitals) and \ref{OPT_deviations} (with optimized orbitals). The upper and lower plots represent N$_2$ at $R=1.0$ \(\text{\AA}\) and $R=1.6$ \(\text{\AA}\), respectively. First, the energy deviations in the upper figures are always smooth, indicating that the errors from the reference-unrelaxed approach at $R=1.0$ \(\text{\AA}\) are insignificant. In contrast, at $R=1.6$ \(\text{\AA}\) with HF orbitals, as shown in Figure \ref{HF_deviations}, the deviation for the fragment containing the $5\sigma_{u}$ secondary orbital, ca. 0.5 mH, is significantly larger than the others. Because of this fragment, all the higher-order fragments that include the $5\sigma_{u}$ orbital have significantly larger deviations, as shown in Figures \ref{HF_deviations}e and \ref{HF_deviations}f. The resulting MBE(2)-UCCGSD/UCCpGSD and MBE(3)-UCCGSD/UCCpGSD energies are 9.6 mH lower and 108.6 mH higher than the corresponding reference-relaxed energies, respectively. In contrast, after employing the optimized orbitals, the energy deviations in Figure \ref{OPT_deviations} become much smaller. The physical reason for this is that the HF orbitals are not well suited for describing multi-reference systems, and the  $5\sigma_{u}$ secondary orbital in the HF solution correlates significantly with the active space, that is, the active space is not well chosen in this particular case. After orbital optimization, the newly defined active space was well-separated from the secondary space; thus, errors from the reference-unrelaxed approximation were significantly suppressed.

It should be noted that the error from the reference-unrelaxed approximation cannot be completely eliminated, even by employing optimized orbitals, and these errors will continue to accumulate as the MBE order increases. As listed in Table \ref{tab:N2_OPT_HF}, the deviation between the reference-unrelaxed and reference-relaxed MBE-UCCGSD/UCCpGSD approaches with optimized orbitals significantly increases at the third order. However, we still prefer to retain this approximation, as it can significantly lower the cost of optimizing the fragment computation. The error caused by the reference-unrelaxed approximation is expected to be significantly suppressed by reasonably enlarging the active space.

\begin{table}[]
	\caption{\label{tab:N2_OPT_HF}
 		Reference-relaxed and reference-unrelaxed MBE-UCCGSD/UCCpGSD energies at each order with the optimized orbitals for N$_2$ at 1.6 \(\text{\AA}\) are listed. The absolute energies are in a.u., and the energy deviations are in mH.}
	\begin{tabular}{cccc}
		\hline
		     & relax & unrelax & $\Delta E$ \\
		\hline\hline
  		MBE(1) & -108.913559 & -108.913420 & 0.139\\
		MBE(2) & -108.957293 & -108.957362 & -0.069\\
  		MBE(3) & -108.958453 & -108.960825 & -2.372 \\
        \hline
	\end{tabular}
\end{table}

\subsection{Vertical Excitation Energy}

After testing the ground-state energies during the bond-breaking process, we now investigate the vertical excitation energies by employing both the MBE-VQE and MBE-VQD algorithms. All calculations presented in this section were performed using the state-averaged optimized orbitals.  

\subsubsection{\rm {LiH}}
The first molecule is the four-electron system LiH. We considered a stretched bond length of 4.8 \r{A} with the cc-pVDZ basis set, for which the FCI results have been reported in Ref. \onlinecite{LiH}. The active space containing the four lowest $\sigma$-orbitals was optimized in the reference calculations. Although there are several dominant determinants in the reference wavefunction ($\ket {\Psi_{\rm ref}}$) of the three lowest $\Sigma^+$ states ($1 ^1\Sigma^+$, $1 ^3\Sigma^+$, $2 ^1\Sigma^+$), this is still a simple system with only two active electrons in a chemical sense. Both UCCSD and UCCGSD can provide highly accurate descriptions of $\ket {\Psi_{\rm ref}}$ and the following MBE-UCCSD and MBE-UCCGSD/UCCpGSD results are satisfactory. As listed in Table \ref{tab:LiH}, they  provide the exact vertical excitation energy, even for the first order. The absolute energies of MBE-UCCGSD/UCCpGSD are almost the same as those of MBE-FCI, and the maximum deviation between MBE-UCCSD and MBE-FCI is 0.007 meV. 

When the HF orbitals are employed without optimization, MBE-UCCGSD/UCCpGSD can still provide accurate excitation energies. However, we found that the description of MBE-UCCSD for $\rm {1 ^3\Sigma^+}$ is problematic; the excitation energies of MBE(1), MBE(2), and MBE(3) orders are 0.08, -0.23 and 0.50 eV, respectively. This poor performance is attributed to the fact that the HF orbitals were optimized for the singlet ground state; hence, the UCCSD operator cannot reasonably describe the triplet excited state. 

\begin{table}[]
	\caption{\label{tab:LiH}
		MBE energies calculated by different approaches for the ground state ($\rm {1 ^1\Sigma^+}$) and two excited states ($\rm {1 ^3\Sigma^+}$, $\rm {2 ^1\Sigma^+}$) of LiH with the cc-PVDZ basis set. The ground state energies are in a.u. and the vertical excitation energies are in eV.}
	\begin{tabular}{cccc}
		\hline
		State & $\rm {1 ^1\Sigma^+}$ & $\rm {1 ^3\Sigma^+}$ & $\rm {2 ^1\Sigma^+}$ \\
		\hline\hline
        \multicolumn{4}{c}{MBE-UCCSD}\\
        \hline
		 \rm {Ref}    & -7.932325 & 0.02 & 1.67 \\
		 \rm {MBE(1)} & -7.932493 & 0.03 & 1.64 \\
		 \rm {MBE(2)} & -7.932521 & 0.03 & 1.64 \\
		 \rm {MBE(3)} & -7.932521 & 0.03 & 1.64 \\
		\hline\hline
        \multicolumn{4}{c}{MBE-UCCGSD/UCCpGSD} \\
        \hline
		 \rm {Ref}    & -7.932329 & 0.02 & 1.67 \\
		 \rm {MBE(1)} & -7.932577 & 0.03 & 1.64 \\
		 \rm {MBE(2)} & -7.932675 & 0.03 & 1.64 \\
		 \rm {MBE(3)} & -7.932667 & 0.03 & 1.64 \\
		\hline\hline
        \multicolumn{4}{c}{MBE-FCI} \\
        \hline
		 \rm {Ref}    & -7.932329 & 0.02 & 1.67 \\
		 \rm {MBE(1)} & -7.932578 & 0.03 & 1.64 \\
		 \rm {MBE(2)} & -7.932671 & 0.03 & 1.64 \\
		 \rm {MBE(3)} & -7.932672 & 0.03 & 1.64 \\
		\hline\hline
         \rm {FCI}    & -7.933503 & 0.03 & 1.64
	\end{tabular}
\end{table}

\subsubsection{\rm {CH$^+$}}
The second molecule is CH$^+$ and we use the same computational conditions as in Section \ref{sec:comparison}. The FCI energies for the three investigated states ($\rm {1 ^1\Sigma^+}$, $\rm {1 ^1\Delta}$, and $\rm {2 ^1\Sigma^+}$) were obtained using the MOLPRO software.\cite{molpro} The ground state ($\rm {1 ^1\Sigma^+}$) is a typical single-reference system, whereas the excited states ($\rm {1 ^1\Delta}$ and $\rm {2 ^1\Sigma^+}$) have two equally dominant doubly-excited determinants but with different signs. The results are tabulated in Table \ref{tab:CH+}. As we have seen, MBE-UCCSD is unable to describe a strong correlation; it can provide reasonable descriptions of the ground state, but performs very poorly for the excited states, particularly in $\rm {1 ^1\Delta}$ of which the reference wavefunction does not have an HF component. In contrast, the MBE-UCCGSD/UCCpGSD absolute energies are much more accurate, and the largest deviation from MBE-FCI is only 0.014 meV. The vertical excitation energies estimated with MBE-UCCGSD/UCCpGSD are almost the same as those estimated with MBE-FCI. This significantly different performance shows that MBE-UCCGSD/UCCpGSD is much more reliable than MBE-UCCSD in describing strongly correlated systems. Even with HF orbitals, the MBE-UCCGSD/UCCpGSD energies are very close to MBE-FCI.
Nevertheless, both converge to FCI rather slowly. The energy deviations of MBE-UCCGSD/UCCpGSD at each order relative to FCI with the optimized and HF orbitals are listed in Table \ref{tab:CH+_HF}. When the optimized orbitals are used, the largest deviations at the second- and third-order are only 2.05 and 0.68 mH, respectively. In contrast, for the HF orbitals, the smallest deviation at the third order is 5.38 mH. This faster convergence rate of MBE as a result of orbital optimization can lower both the computational cost and the error accumulation of MBE effectively.

\begin{table}[]
	\caption{\label{tab:CH+}
		MBE energies calculated by different approaches for the ground state ($\rm {1 ^1\Sigma^+}$) and two excited states ($\rm {1 ^1\Delta}$, $\rm {2 ^1\Sigma^+}$) of CH$^+$ with the aug-cc-pVDZ basis set. The ground state energies are in a.u. and the vertical excitation energies are in eV.}
	\begin{tabular}{cccc}
		\hline
		State & $\rm {1 ^1\Sigma^+}$ & $\rm {1 ^1\Delta}$ & $\rm {2 ^1\Sigma^+}$ \\
		\hline\hline
        \multicolumn{4}{c}{MBE-UCCSD} \\
        \hline
		 \rm {Ref}    & -37.954326 & 7.28 & 9.00 \\
		 \rm {MBE(1)} & -37.992159 & 7.96 & 9.07 \\
		 \rm {MBE(2)} & -38.002705 & 6.73 & 9.91\\
		 \rm {MBE(3)} & -38.002101 &17.93 & 8.20 \\
		\hline\hline
        \multicolumn{4}{c}{MBE-UCCGSD/UCCpGSD} \\
        \hline
		 \rm {Ref}    & -37.954315 & 7.27 & 8.66 \\
		 \rm {MBE(1)} & -37.992902 & 6.86 & 8.58 \\
		 \rm {MBE(2)} & -38.004999 & 6.83 & 8.43 \\
		 \rm {MBE(3)} & -38.003943 & 6.85 & 8.39 \\
		\hline\hline
        \multicolumn{4}{c}{MBE-FCI} \\
        \hline
		 \rm {Ref}    & -37.954315 & 7.27 & 8.66 \\
		 \rm {MBE(1)} & -37.992905 & 6.86 & 8.58 \\
		 \rm {MBE(2)} & -38.004948 & 6.83 & 8.42 \\
		 \rm {MBE(3)} & -38.004445 & 6.86 & 8.40 \\
		\hline\hline
         \rm {FCI}    & -38.004247 & 6.87 & 8.42
	\end{tabular}
\end{table}

\begin{table}[]
	\caption{\label{tab:CH+_HF}
		Energy deviations of MBE-UCCGSD/UCCpGSD relative to FCI with the optimized and HF orbitals. The data are in mH.}
	\begin{tabular}{ccccccc}
		\hline
		State & \multicolumn{2}{c}{$\rm {1 ^1\Sigma^+}$} & \multicolumn{2}{c}{$\rm {1 ^1\Delta}$} & \multicolumn{2}{c}{$\rm {2 ^1\Sigma^+}$} \\
		\hline\hline
             & optimized & HF & optimized & HF & optimized & HF \\
  		\hline
		 \rm {Ref}    & 49.93 & 74.07 & 64.83 & 109.17  & 58.88 & 97.13 \\
		 \rm {MBE(1)} & 11.35 & 16.20 & 11.01 &  11.38  & 17.38 & 34.90 \\
		 \rm {MBE(2)} & -0.75 &  4.57 & -2.05 &   3.24  & -0.52 & 12.89 \\
		 \rm {MBE(3)} &  0.30 &  5.38 & -0.47 &   8.58  & -0.68 &  8.41 \\
		\hline\hline
	\end{tabular}
\end{table}

\subsubsection{\rm {H$_2$O}}
The last system we considered is a water molecule with the aug-cc-pVDZ basis set. The geometry and extrapolated FCI (exFCI) data were obtained from Ref. \onlinecite{H2O}. The O$1s$ orbital was frozen and the next six lowest orbitals were used as the active space with orbital optimization. Similar to the CH$^+$ case, its ground state ($\rm {1 ^1A_1}$) is a typical single-reference system, and the two excited states ($\rm {1 ^3A_1}$ and $\rm {2 ^1A_1}$) have two equally dominant determinants, but with different signs. Here, only the MBE-UCCGSD/UCCpGSD approach was employed. In Table \ref{tab:H2O}, the excitation energy of MBE(3)-UCCGSD/UCCpGSD deviates only by 0.16 eV from that of MBE(3)-FCI, with the largest deviation between their absolute energies being 0.089 meV. Considering that they converged to different targets (UCCGSD and FCI), it is reasonable to assume that more significant deviations occur in larger systems. Because the excitation energy of MBE(3)-UCCGSD/UCCpGSD is only slightly lower than that of FCI and the computational cost increases significantly in the fourth order, truncation of the MBE-UCCGSD/UCCpGSD approach at the third order should be a good compromise for describing the vertical excitation energies in large systems.

\begin{table}[]
	\caption{\label{tab:H2O}
 		Energies of various approaches at different MBE orders for the ground state ($\rm {1 ^1A_1}$) and two excited states ($\rm {1 ^3A_1}$ and $\rm {2 ^1A_1}$) of water with the aug-cc-pVDZ basis set. The ground state energies are in a.u. The vertical excitation energies are in eV.}
	\begin{tabular}{cccc}
		\hline
		State & $\rm {1 ^1A_1}$ & $\rm {1 ^3A_1}$ & $\rm {2 ^1A_1}$ \\
		\hline\hline
        \multicolumn{4}{c} {MBE-UCCGSD/UCCpGSD} \\
        \hline
		 \rm {Ref}    & -76.040609 & 8.47 & 9.03 \\
		 \rm {MBE(1)} & -76.161789 & 9.64 & 9.98 \\
		 \rm {MBE(2)} & -76.288790 & 9.84 & 10.27 \\
		 \rm {MBE(3)} & -76.270582 & 9.34 & 9.66 \\
		\hline\hline
        \multicolumn{4}{c} {MBE-FCI} \\
        \hline
		 \rm {Ref}    & -76.040609 & 8.47 & 9.03 \\
		 \rm {MBE(1)} & -76.161803 & 9.64 & 9.98 \\
		 \rm {MBE(2)} & -76.288923 & 9.83 & 10.25 \\
		 \rm {MBE(3)} & -76.273263 & 9.35 & 9.82 \\
		\hline\hline
         exFCI      & & 9.49 & 9.94
	\end{tabular}
\end{table}

\section{Effect of shot noise}
\begin{figure}[htbp]
\begin{center}
    \includegraphics[width=0.47\textwidth]{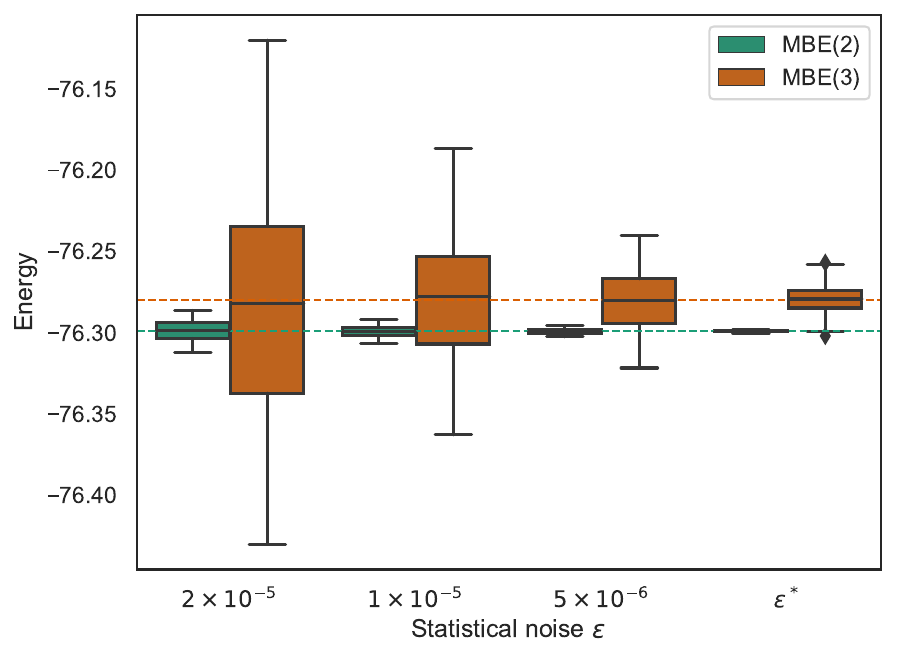}
\end{center}
\caption{Probability plot of the deviation of $E({\rm noise})$ from the exact energy with respect to different $\epsilon$ after 300 times simulations based on the MBE-FCI data of water with the aug-cc-pVDZ basis set.}
\label{probability}
\end{figure}

As we mentioned in Section II, quantum noise, including shot noise, is inevitably present and is included in the total energy. Here, the simulation presented here was therefore designed to quantify the impact of statistical errors caused by shot noise based on the MBE-FCI data for the ground state of water with HF orbitals. Suppose that the noisy energy in fragment $X$ is:
		\begin{equation}
        \begin{aligned}
            e_{X \rm{(noise)}} = e_{X} + \epsilon_X 
        \end{aligned}
		\end{equation}
where $X$ is the composite index that stands for the fragment, $e_X$ is the exact fragment energy, and $\epsilon_X$ represents the corresponding shot noise. In the simulation, $\epsilon_X$ is set as a random number between $\left[ -\epsilon, +\epsilon \right]$, where $\epsilon$ is the controlled maximum statistical error and is a constant. Eq. (10) can be used to calculate all noise increments. The resulting noise MBE(2)-FCI and MBE(3)-FCI energies are denoted as $E_{\rm 2}({\rm noise})$ and $E_{\rm 3}({\rm noise})$, respectively. The exact MBE(2)-FCI (-76.29904) and MBE(3)-FCI (-76.27949) energies were used as accurate references. The simulation was run 300 times with different values of $\epsilon$. The probability plots of the deviations of $E_2({\rm noise})$ and $E_3({\rm noise})$ from the corresponding reference with respect to different $\epsilon$ values are shown in Figure \ref{probability}. In the left three plots in Figure \ref{probability}, $\epsilon$ is fixed at each MBE order, the values of which are set to $2\times10^{-5}$, $1\times10^{-5}$ and $5\times10^{-6}$ Hartree, respectively. In the rightmost plot, we use a varying $\epsilon$, denoted as $\epsilon^{*}$, which is set to $1\times10^{-6}$, $2\times10^{-6}$, $1\times10^{-5}$ and $1\times10^{-4}$ Hartree for the reference and fragment energies at the MBE(1), MBE(2), and MBE(3) orders, respectively. The average $E_2({\rm noise})$ and $E_3({\rm noise})$ (denoted as $\left\langle E_2({\rm noise}) \right\rangle$ and $\left\langle E_3({\rm noise}) \right\rangle$, respectively) and the corresponding standard deviations (denoted as $\sigma2$ and $\sigma3$, respectively) with respect to different $\epsilon$ in each simulation are listed in Table \ref{tab:Noise}. The results in Figure \ref{probability} and Table \ref{tab:Noise} enable us to clearly draw three conclusions: (a), the energy deviations at the third order are significantly larger than those at the second order, attributed to the fact that the error accumulation increases significantly with the MBE order increases; (b), the energy deviations decreased with $\epsilon$. However, the value of $\epsilon$ corresponds to $1/\sqrt{N_{\rm meas}}$, where $N_{\rm meas}$ is the number of measurements, indicating that this is not an efficient way to control noise; (c), in the rightmost plot, the number of measurements corresponding to $\epsilon^{*}$ is actually less than that corresponding to $\epsilon=2\times10^{-5}$ Hartree; however, its energy deviations are much smaller than those of $\epsilon=5\times10^{-6}$ Hartree. This excellent performance demonstrates that the strategy of increasing the number of measurements at lower orders is effective. We strongly recommend the use of different numbers of measurements at different orders in all problem decomposition methods, including MBE.

\begin{table}[]
	\caption{\label{tab:Noise}
 		$\left\langle E_{\rm 2}({\rm noise}) \right\rangle$ and $\left\langle E_{\rm 3}({\rm noise}) \right\rangle$ and the corresponding  $\sigma_2$ and $\sigma_3$ with respect to different $\epsilon$ after 300 times simulations based on the MBE-FCI data of water with the aug-cc-pVDZ basis set are listed. $\left\langle E_{\rm 2}({\rm noise}) \right\rangle$ and $\left\langle E_{\rm 3}({\rm noise}) \right\rangle$ are in a.u.,  $\sigma_2$ and $\sigma_3$ are in mH.}
	\begin{tabular}{ccccc}
		\hline
		  $\epsilon$ & $\left\langle E_{\rm 2}({\rm noise}) \right\rangle$ & $\sigma_2$ & $\left\langle E_{\rm 3}({\rm noise}) \right\rangle$ & $\sigma_3$ \\
		\hline\hline
		  $2\times 10^{-5}$  & -76.29859 & 6.2 & -76.28502 &  68.8 \\
		  $1\times 10^{-5}$    & -76.29917  & 3.3 &-76.27816 &  36.6 \\
		  $5\times 10^{-6}$  & -76.29900  & 1.6 & -76.27995 &  18.1 \\
          $\epsilon^{*}$  & -76.29902 & 0.4 & -76.27938 &  7.8 \\
        \hline
	\end{tabular}
\end{table}

\section{Concluding Remarks}
To deal with the constrained quantum resources in quantum computing, in this study, we combined VQE and VQD with the theoretical framework of MBE. to decompose the tested molecules into fragments by the expansion of secondary orbitals.  Thereupon, we solved the ground state and excited states energies of each fragment by the VQE and VQD algorithms using the (same or different) UCC ansatz. The two approaches we adopted are denoted as MBE-UCCSD and MBE-UCCGSD/UCCpGSD, the latter of which uses the partially generalized UCCGSD operator in the description of the fragment wavefunction and can be considered as an approximation to MBE-UCCGSD. The reference-unrelaxed approximation introduced in this work has demonstrated 
a significant reduction in the parameter-optimization burden during fragment computation. Furthermore, it was shown that the resulting error can be effectively suppressed by employing optimized orbitals.  The ground state energies during the bond-breaking processes of two molecules (H$_2$O and N$_2$) and vertical excitation energies of three small molecules (LiH, CH$^+$ and H$_2$O) were investigated. MBE-UCCSD essentially aims for the UCCSD results with a large basis set and can provide satisfactory descriptions in single-reference systems; however, its performance has been shown to be inferior for multireference systems. In contrast, MBE-UCCGSD/UCCpGSD showed a stable and reliable ability to describe all tests, particularly for multireference systems. The results illustrate that the UCCGSD ansatz is necessary to describe the reference wavefunction which may include a strong static correlation, whereas the partially generalized UCCSD ansatz is sufficiently accurate to capture the dynamical correlation in the fragments. Considering the balance between the rapidly increasing computational cost as the MBE order increases and the accuracy of the results, we recommend using the reference-unrelaxed MBE(3)-UCCGSD/UCCpGSD with optimized orbitals for computing large systems in quantum computing. 

We also designed a simulation to assess the effect of the inevitably present shot noise on the total energy. Our results showed that increasing the number of measurements at the lower orders is a highly efficient way to control the error introduced by shot noise, which can effectively suppress error accumulation without significantly increasing the cost. Finally, we anticipate that our framework could be effectively combined with preprocessing approaches, such as transcorrelated methods for quantum computing, which have been advancing rapidly, to achieve even greater accuracy.

\section*{Associated content}
\subsection*{Supporting information}
The performance of the "reference-relaxed" MBE-UCC methods is shown in SI. The PESs for the bond dissociation processes of H$_2$O and N$_2$ described by MBE-UCCSD, MBE-UCCGSD/UCCpGSD and MRCI+Q are shown in Figures S1 and S2. Furthermore, the deviations of the second- and third-order MBE-UCC energies with respect to MRCI+Q are listed in Tables S1 and S2. Finally, the ground state energies and vertical excitation energies provided by MBE-UCCSD, MBE-UCCGSD/UCCpGSD and MBE-FCI for the three molecules (LiH, CH$^+$ and H$_2$O) are listed in Tables S3, S4 and S5, respectively.

\section*{Acknowledgments}
We wish to thank Siu Chung Tsang for continual computational support. This work was supported by JST, PRESTO (Grant Number JPMJPR2016), Japan. We gratefully acknowledge ECCSE, Kobe University and ACCMS, Kyoto University for their computational resources. 


\begin{mcitethebibliography}{106}
\providecommand*\natexlab[1]{#1}
\providecommand*\mciteSetBstSublistMode[1]{}
\providecommand*\mciteSetBstMaxWidthForm[2]{}
\providecommand*\mciteBstWouldAddEndPuncttrue
  {\def\EndOfBibitem{\unskip.}}
\providecommand*\mciteBstWouldAddEndPunctfalse
  {\let\EndOfBibitem\relax}
\providecommand*\mciteSetBstMidEndSepPunct[3]{}
\providecommand*\mciteSetBstSublistLabelBeginEnd[3]{}
\providecommand*\EndOfBibitem{}
\mciteSetBstSublistMode{f}
\mciteSetBstMaxWidthForm{subitem}{(\alph{mcitesubitemcount})}
\mciteSetBstSublistLabelBeginEnd
  {\mcitemaxwidthsubitemform\space}
  {\relax}
  {\relax}

\bibitem[White(1992)]{DMRG_1}
White,~S.~R. Density matrix formulation for quantum renormalization groups.
  \emph{Phys. Rev. Lett.} \textbf{1992}, \emph{69}, 2863--2866\relax
\mciteBstWouldAddEndPuncttrue
\mciteSetBstMidEndSepPunct{\mcitedefaultmidpunct}
{\mcitedefaultendpunct}{\mcitedefaultseppunct}\relax
\EndOfBibitem
\bibitem[Chan and Sharma(2011)Chan, and Sharma]{DMRG_2}
Chan,~G. K.-L.; Sharma,~S. The Density Matrix Renormalization Group in Quantum
  Chemistry. \emph{Annu. Rev. Phys. Chem.} \textbf{2011}, \emph{62},
  465--481\relax
\mciteBstWouldAddEndPuncttrue
\mciteSetBstMidEndSepPunct{\mcitedefaultmidpunct}
{\mcitedefaultendpunct}{\mcitedefaultseppunct}\relax
\EndOfBibitem
\bibitem[Booth \latin{et~al.}(2009)Booth, Thom, and Alavi]{FCIQMC}
Booth,~G.~H.; Thom,~A. J.~W.; Alavi,~A. {Fermion Monte Carlo without fixed
  nodes: A game of life, death, and annihilation in Slater determinant space}.
  \emph{J. Chem. Phys.} \textbf{2009}, \emph{131}\relax
\mciteBstWouldAddEndPuncttrue
\mciteSetBstMidEndSepPunct{\mcitedefaultmidpunct}
{\mcitedefaultendpunct}{\mcitedefaultseppunct}\relax
\EndOfBibitem
\bibitem[Cleland \latin{et~al.}(2010)Cleland, Booth, and Alavi]{i-FCIQMC}
Cleland,~D.; Booth,~G.~H.; Alavi,~A. {Communications: Survival of the fittest:
  Accelerating convergence in full configuration-interaction quantum Monte
  Carlo}. \emph{J. Chem. Phys.} \textbf{2010}, \emph{132}\relax
\mciteBstWouldAddEndPuncttrue
\mciteSetBstMidEndSepPunct{\mcitedefaultmidpunct}
{\mcitedefaultendpunct}{\mcitedefaultseppunct}\relax
\EndOfBibitem
\bibitem[Sugiyama and Koonin(1986)Sugiyama, and Koonin]{AFQMC}
Sugiyama,~G.; Koonin,~S. Auxiliary field Monte-Carlo for quantum many-body
  ground states. \emph{Ann. Phys.} \textbf{1986}, \emph{168}, 1--26\relax
\mciteBstWouldAddEndPuncttrue
\mciteSetBstMidEndSepPunct{\mcitedefaultmidpunct}
{\mcitedefaultendpunct}{\mcitedefaultseppunct}\relax
\EndOfBibitem
\bibitem[Xu \latin{et~al.}(2018)Xu, Uejima, and Ten-no]{FCCR}
Xu,~E.; Uejima,~M.; Ten-no,~S.~L. Full Coupled-Cluster Reduction for Accurate
  Description of Strong Electron Correlation. \emph{Phys. Rev. Lett.}
  \textbf{2018}, \emph{121}, 113001\relax
\mciteBstWouldAddEndPuncttrue
\mciteSetBstMidEndSepPunct{\mcitedefaultmidpunct}
{\mcitedefaultendpunct}{\mcitedefaultseppunct}\relax
\EndOfBibitem
\bibitem[Xu \latin{et~al.}(2020)Xu, Uejima, and Ten-no]{FCCR2}
Xu,~E.; Uejima,~M.; Ten-no,~S.~L. Towards Near-Exact Solutions of Molecular
  Electronic Structure: Full Coupled-Cluster Reduction with a Second-Order
  Perturbative Correction. \emph{J. Phys. Chem. Lett.} \textbf{2020},
  \emph{11}, 9775--9780\relax
\mciteBstWouldAddEndPuncttrue
\mciteSetBstMidEndSepPunct{\mcitedefaultmidpunct}
{\mcitedefaultendpunct}{\mcitedefaultseppunct}\relax
\EndOfBibitem
\bibitem[Holmes \latin{et~al.}(2016)Holmes, Changlani, and Umrigar]{SHCI}
Holmes,~A.~A.; Changlani,~H.~J.; Umrigar,~C.~J. Efficient Heat-Bath Sampling in
  Fock Space. \emph{J. Phys. Chem. Lett.} \textbf{2016}, \emph{12},
  1561--1571\relax
\mciteBstWouldAddEndPuncttrue
\mciteSetBstMidEndSepPunct{\mcitedefaultmidpunct}
{\mcitedefaultendpunct}{\mcitedefaultseppunct}\relax
\EndOfBibitem
\bibitem[Liu and Hoffmann(2016)Liu, and Hoffmann]{iCI}
Liu,~W.; Hoffmann,~M.~R. iCI: Iterative CI toward full CI. \emph{J. Chem.
  Theory Comput.} \textbf{2016}, \emph{12}, 1169--1178\relax
\mciteBstWouldAddEndPuncttrue
\mciteSetBstMidEndSepPunct{\mcitedefaultmidpunct}
{\mcitedefaultendpunct}{\mcitedefaultseppunct}\relax
\EndOfBibitem
\bibitem[Tubman \latin{et~al.}(2016)Tubman, Lee, Takeshita, Head-Gordon, and
  Whaley]{ASCI}
Tubman,~N.~M.; Lee,~J.; Takeshita,~T.~Y.; Head-Gordon,~M.; Whaley,~K.~B. {A
  deterministic alternative to the full configuration interaction quantum Monte
  Carlo method}. \emph{J. Chem. Phys.} \textbf{2016}, \emph{145}\relax
\mciteBstWouldAddEndPuncttrue
\mciteSetBstMidEndSepPunct{\mcitedefaultmidpunct}
{\mcitedefaultendpunct}{\mcitedefaultseppunct}\relax
\EndOfBibitem
\bibitem[Eriksen \latin{et~al.}(2017)Eriksen, Lipparini, and Gauss]{MBE_1}
Eriksen,~J.~J.; Lipparini,~F.; Gauss,~J. Virtual Orbital Many-Body Expansions:
  A Possible Route towards the Full Configuration Interaction Limit. \emph{J.
  Phys. Chem. Lett.} \textbf{2017}, \emph{8}, 4633\relax
\mciteBstWouldAddEndPuncttrue
\mciteSetBstMidEndSepPunct{\mcitedefaultmidpunct}
{\mcitedefaultendpunct}{\mcitedefaultseppunct}\relax
\EndOfBibitem
\bibitem[Eriksen and Gauss(2019)Eriksen, and Gauss]{MBE_4}
Eriksen,~J.~J.; Gauss,~J. Generalized Many-Body Expanded Full Configuration
  Interaction Theory. \emph{J. Phys. Chem. Lett.} \textbf{2019}, \emph{10},
  7910\relax
\mciteBstWouldAddEndPuncttrue
\mciteSetBstMidEndSepPunct{\mcitedefaultmidpunct}
{\mcitedefaultendpunct}{\mcitedefaultseppunct}\relax
\EndOfBibitem
\bibitem[Manin()]{UseQM_Manin}
Manin,~Y. \emph{Computable and Noncomputable (Sovetskoye Radio, Moscow, 1980),
  pp. 13–15 (in Russian)}\relax
\mciteBstWouldAddEndPuncttrue
\mciteSetBstMidEndSepPunct{\mcitedefaultmidpunct}
{\mcitedefaultendpunct}{\mcitedefaultseppunct}\relax
\EndOfBibitem
\bibitem[Feynman(1982)]{UseQM_Fy}
Feynman,~R.~P. Simulating physics with computers. \emph{Int. J. Theor. Phys.}
  \textbf{1982}, \emph{21}, 467\relax
\mciteBstWouldAddEndPuncttrue
\mciteSetBstMidEndSepPunct{\mcitedefaultmidpunct}
{\mcitedefaultendpunct}{\mcitedefaultseppunct}\relax
\EndOfBibitem
\bibitem[Lloyd(1996)]{QM_simulator}
Lloyd,~S. Universal Quantum Simulators. \emph{Science} \textbf{1996},
  \emph{273}, 1073\relax
\mciteBstWouldAddEndPuncttrue
\mciteSetBstMidEndSepPunct{\mcitedefaultmidpunct}
{\mcitedefaultendpunct}{\mcitedefaultseppunct}\relax
\EndOfBibitem
\bibitem[Aspuru-Guzik \latin{et~al.}(2005)Aspuru-Guzik, Dutoi, Love, and
  Head-Gordon]{PEA_2005}
Aspuru-Guzik,~A.; Dutoi,~A.~D.; Love,~P.~J.; Head-Gordon,~M. Simulated Quantum
  Computation of Molecular Energies. \emph{Science} \textbf{2005}, \emph{309},
  1704\relax
\mciteBstWouldAddEndPuncttrue
\mciteSetBstMidEndSepPunct{\mcitedefaultmidpunct}
{\mcitedefaultendpunct}{\mcitedefaultseppunct}\relax
\EndOfBibitem
\bibitem[Peruzzo \latin{et~al.}(2014)Peruzzo, McClean, Shadbolt, Yung, Zhou,
  Love, Aspuru-Guzik, and O'Brien]{VQE}
Peruzzo,~A.; McClean,~J.; Shadbolt,~P.; Yung,~M.-H.; Zhou,~X.-Q.; Love,~P.~J.;
  Aspuru-Guzik,~A.; O'Brien,~J.~L. A variational eigenvalue solver on a
  photonic quantum processor. \emph{Nat. Commun.} \textbf{2014}, \emph{5},
  4213\relax
\mciteBstWouldAddEndPuncttrue
\mciteSetBstMidEndSepPunct{\mcitedefaultmidpunct}
{\mcitedefaultendpunct}{\mcitedefaultseppunct}\relax
\EndOfBibitem
\bibitem[O'Malley \latin{et~al.}(2016)O'Malley, Babbush, Kivlichan, Romero,
  McClean, Barends, Kelly, Roushan, Tranter, Ding, Campbell, Chen, Chen,
  Chiaro, Dunsworth, Fowler, Jeffrey, Lucero, Megrant, Mutus, Neeley, Neill,
  Quintana, Sank, Vainsencher, Wenner, White, Coveney, Love, Neven,
  Aspuru-Guzik, and Martinis]{QM_energy_2016}
O'Malley,~P. J.~J.; Babbush,~R.; Kivlichan,~I.~D.; Romero,~J.; McClean,~J.~R.;
  Barends,~R.; Kelly,~J.; Roushan,~P.; Tranter,~A.; Ding,~N.; Campbell,~B.;
  Chen,~Y.; Chen,~Z.; Chiaro,~B.; Dunsworth,~A.; Fowler,~A.~G.; Jeffrey,~E.;
  Lucero,~E.; Megrant,~A.; Mutus,~J.~Y.; Neeley,~M.; Neill,~C.; Quintana,~C.;
  Sank,~D.; Vainsencher,~A.; Wenner,~J.; White,~T.~C.; Coveney,~P.~V.;
  Love,~P.~J.; Neven,~H.; Aspuru-Guzik,~A.; Martinis,~J.~M. Scalable Quantum
  Simulation of Molecular Energies. \emph{Phys. Rev. X} \textbf{2016},
  \emph{6}, 031007\relax
\mciteBstWouldAddEndPuncttrue
\mciteSetBstMidEndSepPunct{\mcitedefaultmidpunct}
{\mcitedefaultendpunct}{\mcitedefaultseppunct}\relax
\EndOfBibitem
\bibitem[Kandala \latin{et~al.}(2017)Kandala, Mezzacapo, Temme, Takita, Brink,
  Chow, and Gambetta]{QM_energy_2017}
Kandala,~A.; Mezzacapo,~A.; Temme,~K.; Takita,~M.; Brink,~M.; Chow,~J.~M.;
  Gambetta,~J.~M. Hardware-efficient variational quantum eigensolver for small
  molecules and quantum magnets. \emph{Nature} \textbf{2017}, \emph{549},
  242\relax
\mciteBstWouldAddEndPuncttrue
\mciteSetBstMidEndSepPunct{\mcitedefaultmidpunct}
{\mcitedefaultendpunct}{\mcitedefaultseppunct}\relax
\EndOfBibitem
\bibitem[Hempel \latin{et~al.}(2018)Hempel, Maier, Romero, McClean, Monz, Shen,
  Jurcevic, Lanyon, Love, Babbush, Aspuru-Guzik, Blatt, and
  Roos]{QM_energy_2018}
Hempel,~C.; Maier,~C.; Romero,~J.; McClean,~J.; Monz,~T.; Shen,~H.;
  Jurcevic,~P.; Lanyon,~B.~P.; Love,~P.; Babbush,~R.; Aspuru-Guzik,~A.;
  Blatt,~R.; Roos,~C.~F. Quantum Chemistry Calculations on a Trapped-Ion
  Quantum Simulator. \emph{Phys. Rev. X} \textbf{2018}, \emph{8}, 031022\relax
\mciteBstWouldAddEndPuncttrue
\mciteSetBstMidEndSepPunct{\mcitedefaultmidpunct}
{\mcitedefaultendpunct}{\mcitedefaultseppunct}\relax
\EndOfBibitem
\bibitem[Fowler \latin{et~al.}(2012)Fowler, Mariantoni, Martinis, and
  Cleland]{FTQC_1}
Fowler,~A.~G.; Mariantoni,~M.; Martinis,~J.~M.; Cleland,~A.~N. Surface codes:
  Towards practical large-scale quantum computation. \emph{Phys. Rev. A}
  \textbf{2012}, \emph{86}, 032324\relax
\mciteBstWouldAddEndPuncttrue
\mciteSetBstMidEndSepPunct{\mcitedefaultmidpunct}
{\mcitedefaultendpunct}{\mcitedefaultseppunct}\relax
\EndOfBibitem
\bibitem[Yoder \latin{et~al.}(2016)Yoder, Takagi, and Chuang]{FTQC_2}
Yoder,~T.~J.; Takagi,~R.; Chuang,~I.~L. Universal Fault-Tolerant Gates on
  Concatenated Stabilizer Codes. \emph{Phys. Rev. X} \textbf{2016}, \emph{6},
  031039\relax
\mciteBstWouldAddEndPuncttrue
\mciteSetBstMidEndSepPunct{\mcitedefaultmidpunct}
{\mcitedefaultendpunct}{\mcitedefaultseppunct}\relax
\EndOfBibitem
\bibitem[Fowler and Gidney(2019)Fowler, and Gidney]{FTQC_3}
Fowler,~A.~G.; Gidney,~C. Low overhead quantum computation using lattice
  surgery. \textbf{2019}, arXiv:1808.06709 [quant--ph], date of access:
  2023/9/29\relax
\mciteBstWouldAddEndPuncttrue
\mciteSetBstMidEndSepPunct{\mcitedefaultmidpunct}
{\mcitedefaultendpunct}{\mcitedefaultseppunct}\relax
\EndOfBibitem
\bibitem[Suzuki \latin{et~al.}(2022)Suzuki, Endo, Fujii, and Tokunaga]{FTQC_4}
Suzuki,~Y.; Endo,~S.; Fujii,~K.; Tokunaga,~Y. Quantum Error Mitigation as a
  Universal Error Reduction Technique: Applications from the NISQ to the
  Fault-Tolerant Quantum Computing Eras. \emph{PRX Quantum} \textbf{2022},
  \emph{3}, 010345\relax
\mciteBstWouldAddEndPuncttrue
\mciteSetBstMidEndSepPunct{\mcitedefaultmidpunct}
{\mcitedefaultendpunct}{\mcitedefaultseppunct}\relax
\EndOfBibitem
\bibitem[Ganzhorn \latin{et~al.}(2019)Ganzhorn, Egger, Barkoutsos, Ollitrault,
  Salis, Moll, Roth, Fuhrer, Mueller, Woerner, Tavernelli, and
  Filipp]{classical_QM_1}
Ganzhorn,~M.; Egger,~D.; Barkoutsos,~P.; Ollitrault,~P.; Salis,~G.; Moll,~N.;
  Roth,~M.; Fuhrer,~A.; Mueller,~P.; Woerner,~S.; Tavernelli,~I.; Filipp,~S.
  Gate-Efficient Simulation of Molecular Eigenstates on a Quantum Computer.
  \emph{Phys. Rev. Appl.} \textbf{2019}, \emph{11}, 044092\relax
\mciteBstWouldAddEndPuncttrue
\mciteSetBstMidEndSepPunct{\mcitedefaultmidpunct}
{\mcitedefaultendpunct}{\mcitedefaultseppunct}\relax
\EndOfBibitem
\bibitem[Nakanishi \latin{et~al.}(2019)Nakanishi, Mitarai, and
  Fujii]{classical_QM_2}
Nakanishi,~K.~M.; Mitarai,~K.; Fujii,~K. Subspace-search variational quantum
  eigensolver for excited states. \emph{Phys. Rev. Res.} \textbf{2019},
  \emph{1}, 033062\relax
\mciteBstWouldAddEndPuncttrue
\mciteSetBstMidEndSepPunct{\mcitedefaultmidpunct}
{\mcitedefaultendpunct}{\mcitedefaultseppunct}\relax
\EndOfBibitem
\bibitem[Babbush \latin{et~al.}(2018)Babbush, Wiebe, McClean, McClain, Neven,
  and Chan]{Low_depth_PRX}
Babbush,~R.; Wiebe,~N.; McClean,~J.; McClain,~J.; Neven,~H.; Chan,~G. K.-L.
  Low-Depth Quantum Simulation of Materials. \emph{Phys. Rev. X} \textbf{2018},
  \emph{8}, 011044\relax
\mciteBstWouldAddEndPuncttrue
\mciteSetBstMidEndSepPunct{\mcitedefaultmidpunct}
{\mcitedefaultendpunct}{\mcitedefaultseppunct}\relax
\EndOfBibitem
\bibitem[Kivlichan \latin{et~al.}(2018)Kivlichan, McClean, Wiebe, Gidney,
  Aspuru-Guzik, Chan, and Babbush]{Low_depth_PRL}
Kivlichan,~I.~D.; McClean,~J.; Wiebe,~N.; Gidney,~C.; Aspuru-Guzik,~A.;
  Chan,~G. K.-L.; Babbush,~R. Quantum Simulation of Electronic Structure with
  Linear Depth and Connectivity. \emph{Phys. Rev. Lett.} \textbf{2018},
  \emph{120}, 110501\relax
\mciteBstWouldAddEndPuncttrue
\mciteSetBstMidEndSepPunct{\mcitedefaultmidpunct}
{\mcitedefaultendpunct}{\mcitedefaultseppunct}\relax
\EndOfBibitem
\bibitem[Ryabinkin \latin{et~al.}(2018)Ryabinkin, Yen, Genin, and
  Izmaylov]{Low_depth_JCTC}
Ryabinkin,~I.~G.; Yen,~T.-C.; Genin,~S.~N.; Izmaylov,~A.~F. Qubit Coupled
  Cluster Method: A Systematic Approach to Quantum Chemistry on a Quantum
  Computer. \emph{J. Chem. Theory Comput.} \textbf{2018}, \emph{14}, 6317\relax
\mciteBstWouldAddEndPuncttrue
\mciteSetBstMidEndSepPunct{\mcitedefaultmidpunct}
{\mcitedefaultendpunct}{\mcitedefaultseppunct}\relax
\EndOfBibitem
\bibitem[Grimsley \latin{et~al.}(2019)Grimsley, Economou, Barnes, and
  Mayhall]{Low_depth_NC}
Grimsley,~H.~R.; Economou,~S.~E.; Barnes,~E.; Mayhall,~N.~J. An adaptive
  variational algorithm for exact molecular simulations on a quantum computer.
  \emph{Nat. Commun.} \textbf{2019}, \emph{10}, 3007\relax
\mciteBstWouldAddEndPuncttrue
\mciteSetBstMidEndSepPunct{\mcitedefaultmidpunct}
{\mcitedefaultendpunct}{\mcitedefaultseppunct}\relax
\EndOfBibitem
\bibitem[Tsuchimochi \latin{et~al.}(2020)Tsuchimochi, Mori, and
  Ten-no]{Tsuchimochi20}
Tsuchimochi,~T.; Mori,~Y.; Ten-no,~S.~L. Spin-projection for quantum
  computation: A low-depth approach to strong correlation. \emph{Phys. Rev.
  Research} \textbf{2020}, \emph{2}, 043142\relax
\mciteBstWouldAddEndPuncttrue
\mciteSetBstMidEndSepPunct{\mcitedefaultmidpunct}
{\mcitedefaultendpunct}{\mcitedefaultseppunct}\relax
\EndOfBibitem
\bibitem[Tsuchimochi \latin{et~al.}(2022)Tsuchimochi, Taii, Nishimaki, and
  Ten-no]{Tsuchimochi22}
Tsuchimochi,~T.; Taii,~M.; Nishimaki,~T.; Ten-no,~S.~L. Adaptive construction
  of shallower quantum circuits with quantum spin projection for fermionic
  systems. \emph{Phys. Rev. Research} \textbf{2022}, \emph{4}, 033100\relax
\mciteBstWouldAddEndPuncttrue
\mciteSetBstMidEndSepPunct{\mcitedefaultmidpunct}
{\mcitedefaultendpunct}{\mcitedefaultseppunct}\relax
\EndOfBibitem
\bibitem[Bauman \latin{et~al.}(2019)Bauman, Bylaska, Krishnamoorthy, Low,
  Wiebe, Granade, Roetteler, Troyer, and Kowalski]{Bauman19A}
Bauman,~N.~P.; Bylaska,~E.~J.; Krishnamoorthy,~S.; Low,~G.~H.; Wiebe,~N.;
  Granade,~C.~E.; Roetteler,~M.; Troyer,~M.; Kowalski,~K. Downfolding of
  many-body Hamiltonians using active-space models: Extension of the sub-system
  embedding sub-algebras approach to unitary coupled cluster formalisms.
  \emph{J. Chem. Phys.} \textbf{2019}, \emph{151}, 014107\relax
\mciteBstWouldAddEndPuncttrue
\mciteSetBstMidEndSepPunct{\mcitedefaultmidpunct}
{\mcitedefaultendpunct}{\mcitedefaultseppunct}\relax
\EndOfBibitem
\bibitem[Bauman \latin{et~al.}(2019)Bauman, Low, and Kowalski]{Bauman19B}
Bauman,~N.~P.; Low,~G.~H.; Kowalski,~K. Quantum simulations of excited states
  with active-space downfolded Hamiltonians. \emph{J. Chem. Phys.}
  \textbf{2019}, \emph{151}, 234114\relax
\mciteBstWouldAddEndPuncttrue
\mciteSetBstMidEndSepPunct{\mcitedefaultmidpunct}
{\mcitedefaultendpunct}{\mcitedefaultseppunct}\relax
\EndOfBibitem
\bibitem[McArdle and Tew(2020)McArdle, and Tew]{mcardle2020improving}
McArdle,~S.; Tew,~D.~P. Improving the accuracy of quantum computational
  chemistry using the transcorrelated method. \textbf{2020}, arXiv:2006.11181
  [quant--ph], date of access: 2023/9/29\relax
\mciteBstWouldAddEndPuncttrue
\mciteSetBstMidEndSepPunct{\mcitedefaultmidpunct}
{\mcitedefaultendpunct}{\mcitedefaultseppunct}\relax
\EndOfBibitem
\bibitem[Sokolov \latin{et~al.}(2023)Sokolov, Dobrautz, Luo, Alavi, and
  Tavernelli]{sokolov2023orders}
Sokolov,~I.~O.; Dobrautz,~W.; Luo,~H.; Alavi,~A.; Tavernelli,~I. Orders of
  magnitude reduction in the computational overhead for quantum many-body
  problems on quantum computers via an exact transcorrelated method.
  \textbf{2023}, arXiv:2201.03049 [quant--ph], date of access: 2023/9/29\relax
\mciteBstWouldAddEndPuncttrue
\mciteSetBstMidEndSepPunct{\mcitedefaultmidpunct}
{\mcitedefaultendpunct}{\mcitedefaultseppunct}\relax
\EndOfBibitem
\bibitem[Ten-no(2023)]{Tenno2023}
Ten-no,~S.~L. Nonunitary projective transcorrelation theory inspired by the F12
  ansatz. \emph{J. Chem. Phys.} \textbf{2023}, in press.\relax
\mciteBstWouldAddEndPunctfalse
\mciteSetBstMidEndSepPunct{\mcitedefaultmidpunct}
{}{\mcitedefaultseppunct}\relax
\EndOfBibitem
\bibitem[Yanai and Shiozaki(2012)Yanai, and Shiozaki]{Yanai2012}
Yanai,~T.; Shiozaki,~T. {Canonical transcorrelated theory with projected
  Slater-type geminals}. \emph{The Journal of Chemical Physics} \textbf{2012},
  \emph{136}, 084107\relax
\mciteBstWouldAddEndPuncttrue
\mciteSetBstMidEndSepPunct{\mcitedefaultmidpunct}
{\mcitedefaultendpunct}{\mcitedefaultseppunct}\relax
\EndOfBibitem
\bibitem[Kumar \latin{et~al.}(2022)Kumar, Asthana, Masteran, Valeev, Zhang,
  Cincio, Tretiak, and Dub]{Kumar2022}
Kumar,~A.; Asthana,~A.; Masteran,~C.; Valeev,~E.~F.; Zhang,~Y.; Cincio,~L.;
  Tretiak,~S.; Dub,~P.~A. Quantum Simulation of Molecular Electronic States
  with a Transcorrelated Hamiltonian: Higher Accuracy with Fewer Qubits.
  \emph{J. Chem. Theory Comput.} \textbf{2022}, \emph{18}, 5312--5324\relax
\mciteBstWouldAddEndPuncttrue
\mciteSetBstMidEndSepPunct{\mcitedefaultmidpunct}
{\mcitedefaultendpunct}{\mcitedefaultseppunct}\relax
\EndOfBibitem
\bibitem[Gordon \latin{et~al.}(2012)Gordon, Fedorov, Pruitt, and
  Slipchenko]{PD_rev1}
Gordon,~M.~S.; Fedorov,~D.~G.; Pruitt,~S.~R.; Slipchenko,~L.~V. Fragmentation
  Methods: A Route to Accurate Calculations on Large Systems. \emph{Chem. Rev.}
  \textbf{2012}, \emph{112}, 632\relax
\mciteBstWouldAddEndPuncttrue
\mciteSetBstMidEndSepPunct{\mcitedefaultmidpunct}
{\mcitedefaultendpunct}{\mcitedefaultseppunct}\relax
\EndOfBibitem
\bibitem[Collins and Bettens(2015)Collins, and Bettens]{PD_rev2}
Collins,~M.~A.; Bettens,~R. P.~A. Energy-Based Molecular Fragmentation Methods.
  \emph{Chem. Rev.} \textbf{2015}, \emph{115}, 5607\relax
\mciteBstWouldAddEndPuncttrue
\mciteSetBstMidEndSepPunct{\mcitedefaultmidpunct}
{\mcitedefaultendpunct}{\mcitedefaultseppunct}\relax
\EndOfBibitem
\bibitem[Raghavachari and Saha(2015)Raghavachari, and Saha]{PD_rev3}
Raghavachari,~K.; Saha,~A. Accurate Composite and Fragment-Based Quantum
  Chemical Models for Large Molecules. \emph{Chem. Rev.} \textbf{2015},
  \emph{115}, 5643\relax
\mciteBstWouldAddEndPuncttrue
\mciteSetBstMidEndSepPunct{\mcitedefaultmidpunct}
{\mcitedefaultendpunct}{\mcitedefaultseppunct}\relax
\EndOfBibitem
\bibitem[Sun and Chan(2016)Sun, and Chan]{PD_rev4}
Sun,~Q.; Chan,~G. K.-L. Quantum Embedding Theories. \emph{Acc. Chem. Res.}
  \textbf{2016}, \emph{49}, 2705\relax
\mciteBstWouldAddEndPuncttrue
\mciteSetBstMidEndSepPunct{\mcitedefaultmidpunct}
{\mcitedefaultendpunct}{\mcitedefaultseppunct}\relax
\EndOfBibitem
\bibitem[Verma \latin{et~al.}(2021)Verma, Huntington, Coons, Kawashima,
  Yamazaki, and Zaribafiyan]{Verma21}
Verma,~P.; Huntington,~L.; Coons,~M.~P.; Kawashima,~Y.; Yamazaki,~T.;
  Zaribafiyan,~A. Scaling up electronic structure calculations on quantum
  computers: The frozen natural orbital based method of increments. \emph{J.
  Chem. Phys.} \textbf{2021}, \emph{155}, 034110\relax
\mciteBstWouldAddEndPuncttrue
\mciteSetBstMidEndSepPunct{\mcitedefaultmidpunct}
{\mcitedefaultendpunct}{\mcitedefaultseppunct}\relax
\EndOfBibitem
\bibitem[Fujii \latin{et~al.}(2022)Fujii, Mizuta, Ueda, Mitarai, Mizukami, and
  Nakagawa]{Fujii2022}
Fujii,~K.; Mizuta,~K.; Ueda,~H.; Mitarai,~K.; Mizukami,~W.; Nakagawa,~Y.~O.
  Deep Variational Quantum Eigensolver: A Divide-And-Conquer Method for Solving
  a Larger Problem with Smaller Size Quantum Computers. \emph{PRX Quantum}
  \textbf{2022}, \emph{3}, 010346\relax
\mciteBstWouldAddEndPuncttrue
\mciteSetBstMidEndSepPunct{\mcitedefaultmidpunct}
{\mcitedefaultendpunct}{\mcitedefaultseppunct}\relax
\EndOfBibitem
\bibitem[Yoshikawa \latin{et~al.}(2022)Yoshikawa, Takanashi, and
  Nakai]{Yoshikawa2022}
Yoshikawa,~T.; Takanashi,~T.; Nakai,~H. Quantum Algorithm of the
  Divide-and-Conquer Unitary Coupled Cluster Method with a Variational Quantum
  Eigensolver. \emph{Journal of Chemical Theory and Computation} \textbf{2022},
  \emph{18}, 5360--5373, PMID: 35926142\relax
\mciteBstWouldAddEndPuncttrue
\mciteSetBstMidEndSepPunct{\mcitedefaultmidpunct}
{\mcitedefaultendpunct}{\mcitedefaultseppunct}\relax
\EndOfBibitem
\bibitem[Bethe and Goldstone(1957)Bethe, and Goldstone]{BG_equation}
Bethe,~H.~A.; Goldstone,~J. Effect of a Repulsive Core in the Theory of Complex
  Nuclei. \emph{Proc. R. Soc. A} \textbf{1957}, \emph{238}, 551\relax
\mciteBstWouldAddEndPuncttrue
\mciteSetBstMidEndSepPunct{\mcitedefaultmidpunct}
{\mcitedefaultendpunct}{\mcitedefaultseppunct}\relax
\EndOfBibitem
\bibitem[Nesbet(1967)]{BG_equation_N1}
Nesbet,~R.~K. Atomic Bethe-Goldstone Equations. I. The Be Atom. \emph{Phys.
  Rev.} \textbf{1967}, \emph{155}, 51\relax
\mciteBstWouldAddEndPuncttrue
\mciteSetBstMidEndSepPunct{\mcitedefaultmidpunct}
{\mcitedefaultendpunct}{\mcitedefaultseppunct}\relax
\EndOfBibitem
\bibitem[Nesbet(1967)]{BG_equation_N2}
Nesbet,~R.~K. Atomic Bethe-Goldstone Equations. II. The Ne Atom. \emph{Phys.
  Rev.} \textbf{1967}, \emph{155}, 56\relax
\mciteBstWouldAddEndPuncttrue
\mciteSetBstMidEndSepPunct{\mcitedefaultmidpunct}
{\mcitedefaultendpunct}{\mcitedefaultseppunct}\relax
\EndOfBibitem
\bibitem[Nesbet(1968)]{BG_equation_N3}
Nesbet,~R.~K. Atomic Bethe-Goldstone Equations. III. Correlation Energies of
  Ground States of Be, B, C, N, O, F, and Ne. \emph{Phys. Rev.} \textbf{1968},
  \emph{175}, 2\relax
\mciteBstWouldAddEndPuncttrue
\mciteSetBstMidEndSepPunct{\mcitedefaultmidpunct}
{\mcitedefaultendpunct}{\mcitedefaultseppunct}\relax
\EndOfBibitem
\bibitem[Zimmerman(2017)]{iFCI_1}
Zimmerman,~P.~M. Incremental full configuration interaction. \emph{J. Chem.
  Phys.} \textbf{2017}, \emph{146}, 104102\relax
\mciteBstWouldAddEndPuncttrue
\mciteSetBstMidEndSepPunct{\mcitedefaultmidpunct}
{\mcitedefaultendpunct}{\mcitedefaultseppunct}\relax
\EndOfBibitem
\bibitem[Zimmerman(2017)]{iFCI_2}
Zimmerman,~P.~M. Singlet–Triplet Gaps through Incremental Full Configuration
  Interaction. \emph{J. Phys. Chem. A} \textbf{2017}, \emph{121}, 4712\relax
\mciteBstWouldAddEndPuncttrue
\mciteSetBstMidEndSepPunct{\mcitedefaultmidpunct}
{\mcitedefaultendpunct}{\mcitedefaultseppunct}\relax
\EndOfBibitem
\bibitem[Zimmerman(2017)]{iFCI_3}
Zimmerman,~P.~M. Strong correlation in incremental full configuration
  interaction. \emph{J. Chem. Phys.} \textbf{2017}, \emph{146}, 224104\relax
\mciteBstWouldAddEndPuncttrue
\mciteSetBstMidEndSepPunct{\mcitedefaultmidpunct}
{\mcitedefaultendpunct}{\mcitedefaultseppunct}\relax
\EndOfBibitem
\bibitem[Eriksen and Gauss(2018)Eriksen, and Gauss]{MBE_2}
Eriksen,~J.~J.; Gauss,~J. Many-Body Expanded Full Configuration Interaction. I.
  Weakly Correlated Regime. \emph{J. Chem. Theory Comput.} \textbf{2018},
  \emph{14}, 5180\relax
\mciteBstWouldAddEndPuncttrue
\mciteSetBstMidEndSepPunct{\mcitedefaultmidpunct}
{\mcitedefaultendpunct}{\mcitedefaultseppunct}\relax
\EndOfBibitem
\bibitem[Eriksen and Gauss(2019)Eriksen, and Gauss]{MBE_3}
Eriksen,~J.~J.; Gauss,~J. Many-Body Expanded Full Configuration Interaction.
  II. Strongly Correlated Regime. \emph{J. Chem. Theory Comput.} \textbf{2019},
  \emph{15}, 4873\relax
\mciteBstWouldAddEndPuncttrue
\mciteSetBstMidEndSepPunct{\mcitedefaultmidpunct}
{\mcitedefaultendpunct}{\mcitedefaultseppunct}\relax
\EndOfBibitem
\bibitem[Romero \latin{et~al.}(2018)Romero, Babbush, McClean, Hempel, Love, and
  Aspuru-Guzik]{Trotter_3}
Romero,~J.; Babbush,~R.; McClean,~J.~R.; Hempel,~C.; Love,~P.~J.;
  Aspuru-Guzik,~A. Strategies for quantum computing molecular energies using
  the unitary coupled cluster ansatz. \emph{Quantum Sci. Technol.}
  \textbf{2018}, \emph{4}, 014008\relax
\mciteBstWouldAddEndPuncttrue
\mciteSetBstMidEndSepPunct{\mcitedefaultmidpunct}
{\mcitedefaultendpunct}{\mcitedefaultseppunct}\relax
\EndOfBibitem
\bibitem[K{\"u}hn \latin{et~al.}(2019)K{\"u}hn, Zanker, Deglmann, Marthaler,
  and Wei{\ss}]{UCCSD_VQE}
K{\"u}hn,~M.; Zanker,~S.; Deglmann,~P.; Marthaler,~M.; Wei{\ss},~H. Accuracy
  and Resource Estimations for Quantum Chemistry on a Near-Term Quantum
  Computer. \emph{J. Chem. Theory Comput.} \textbf{2019}, \emph{15}, 4764\relax
\mciteBstWouldAddEndPuncttrue
\mciteSetBstMidEndSepPunct{\mcitedefaultmidpunct}
{\mcitedefaultendpunct}{\mcitedefaultseppunct}\relax
\EndOfBibitem
\bibitem[Yoshikawa \latin{et~al.}(2022)Yoshikawa, Takanashi, and
  Nakai]{DC_qUCC_VQE}
Yoshikawa,~T.; Takanashi,~T.; Nakai,~H. Quantum Algorithm of the
  Divide-and-Conquer Unitary Coupled Cluster Method with a Variational Quantum
  Eigensolver. \emph{J. Chem. Theory Comput.} \textbf{2022}, \emph{18},
  5360--5373\relax
\mciteBstWouldAddEndPuncttrue
\mciteSetBstMidEndSepPunct{\mcitedefaultmidpunct}
{\mcitedefaultendpunct}{\mcitedefaultseppunct}\relax
\EndOfBibitem
\bibitem[Higgott \latin{et~al.}(2019)Higgott, Wang, and Brierley]{VQD}
Higgott,~O.; Wang,~D.; Brierley,~S. Variational Quantum Computation of Excited
  States. \emph{Quantum} \textbf{2019}, \emph{3}, 156\relax
\mciteBstWouldAddEndPuncttrue
\mciteSetBstMidEndSepPunct{\mcitedefaultmidpunct}
{\mcitedefaultendpunct}{\mcitedefaultseppunct}\relax
\EndOfBibitem
\bibitem[Eriksen and Gauss(2020)Eriksen, and Gauss]{MBE_5}
Eriksen,~J.~J.; Gauss,~J. Ground and Excited State First-Order Properties in
  Many Body Expanded Full Configuration Interaction Theory. \emph{J. Chem.
  Phys.} \textbf{2020}, \emph{153}, 154107\relax
\mciteBstWouldAddEndPuncttrue
\mciteSetBstMidEndSepPunct{\mcitedefaultmidpunct}
{\mcitedefaultendpunct}{\mcitedefaultseppunct}\relax
\EndOfBibitem
\bibitem[Eriksen(2021)]{MBE_6}
Eriksen,~J.~J. The Shape of Full Configuration Interaction to Come. \emph{J.
  Phys. Chem. Lett.} \textbf{2021}, \emph{12}, 418\relax
\mciteBstWouldAddEndPuncttrue
\mciteSetBstMidEndSepPunct{\mcitedefaultmidpunct}
{\mcitedefaultendpunct}{\mcitedefaultseppunct}\relax
\EndOfBibitem
\bibitem[Eriksen and Gauss(2021)Eriksen, and Gauss]{MBE_7}
Eriksen,~J.~J.; Gauss,~J. Incremental treatments of the full configuration
  interaction problem. \emph{WIREs Comput Mol Sci.} \textbf{2021}, \emph{11},
  e1525\relax
\mciteBstWouldAddEndPuncttrue
\mciteSetBstMidEndSepPunct{\mcitedefaultmidpunct}
{\mcitedefaultendpunct}{\mcitedefaultseppunct}\relax
\EndOfBibitem
\bibitem[Nooijen(2000)]{CCGSD_Nooijen}
Nooijen,~M. Can the Eigenstates of a Many-Body Hamiltonian Be Represented
  Exactly Using a General Two-Body Cluster Expansion? \emph{Phys. Rev. Lett.}
  \textbf{2000}, \emph{84}, 2108--2111\relax
\mciteBstWouldAddEndPuncttrue
\mciteSetBstMidEndSepPunct{\mcitedefaultmidpunct}
{\mcitedefaultendpunct}{\mcitedefaultseppunct}\relax
\EndOfBibitem
\bibitem[Nakatsuji(2000)]{CCGSD_Nakatsuji}
Nakatsuji,~H. Structure of the exact wave function. \emph{J. Chem. Phys.}
  \textbf{2000}, \emph{113}, 2949--2956\relax
\mciteBstWouldAddEndPuncttrue
\mciteSetBstMidEndSepPunct{\mcitedefaultmidpunct}
{\mcitedefaultendpunct}{\mcitedefaultseppunct}\relax
\EndOfBibitem
\bibitem[Van~Voorhis and Head-Gordon(2001)Van~Voorhis, and
  Head-Gordon]{CCGSD_Troy}
Van~Voorhis,~T.; Head-Gordon,~M. Two-body coupled cluster expansions. \emph{J.
  Chem. Phys.} \textbf{2001}, \emph{115}, 5033--5040\relax
\mciteBstWouldAddEndPuncttrue
\mciteSetBstMidEndSepPunct{\mcitedefaultmidpunct}
{\mcitedefaultendpunct}{\mcitedefaultseppunct}\relax
\EndOfBibitem
\bibitem[Piecuch \latin{et~al.}(2003)Piecuch, Kowalski, Fan, and
  Jedziniak]{CCGSD_Piecuch}
Piecuch,~P.; Kowalski,~K.; Fan,~P.-D.; Jedziniak,~K. Exactness of Two-Body
  Cluster Expansions in Many-Body Quantum Theory. \emph{Phys. Rev. Lett.}
  \textbf{2003}, \emph{90}, 113001\relax
\mciteBstWouldAddEndPuncttrue
\mciteSetBstMidEndSepPunct{\mcitedefaultmidpunct}
{\mcitedefaultendpunct}{\mcitedefaultseppunct}\relax
\EndOfBibitem
\bibitem[Davidson(2003)]{CCGSD_Davidson}
Davidson,~E.~R. Exactness of the General Two-Body Cluster Expansion in
  Many-Body Quantum Theory. \emph{Phys. Rev. Lett.} \textbf{2003}, \emph{91},
  123001\relax
\mciteBstWouldAddEndPuncttrue
\mciteSetBstMidEndSepPunct{\mcitedefaultmidpunct}
{\mcitedefaultendpunct}{\mcitedefaultseppunct}\relax
\EndOfBibitem
\bibitem[Ronen(2003)]{CCGSD_Nonen}
Ronen,~S. Can the Eigenstates of a Many-Body Hamiltonian Be Represented Exactly
  Using a General Two-Body Cluster Expansion? \emph{Phys. Rev. Lett.}
  \textbf{2003}, \emph{91}, 123002\relax
\mciteBstWouldAddEndPuncttrue
\mciteSetBstMidEndSepPunct{\mcitedefaultmidpunct}
{\mcitedefaultendpunct}{\mcitedefaultseppunct}\relax
\EndOfBibitem
\bibitem[Mukherjee and Kutzelnigg(2004)Mukherjee, and
  Kutzelnigg]{CCGSD_Mukherjee}
Mukherjee,~D.; Kutzelnigg,~W. Some comments on the coupled cluster with
  generalized singles and doubles (CCGSD) ansatz. \emph{Chem. Phys. Lett.}
  \textbf{2004}, \emph{397}, 174--179\relax
\mciteBstWouldAddEndPuncttrue
\mciteSetBstMidEndSepPunct{\mcitedefaultmidpunct}
{\mcitedefaultendpunct}{\mcitedefaultseppunct}\relax
\EndOfBibitem
\bibitem[Lee \latin{et~al.}(2019)Lee, Huggins, Head-Gordon, and Whaley]{UCCGSD}
Lee,~J.; Huggins,~W.~J.; Head-Gordon,~M.; Whaley,~K.~B. Generalized Unitary
  Coupled Cluster Wave functions for Quantum Computation. \emph{J. Chem. Theory
  Comput.} \textbf{2019}, \emph{15}, 311--324\relax
\mciteBstWouldAddEndPuncttrue
\mciteSetBstMidEndSepPunct{\mcitedefaultmidpunct}
{\mcitedefaultendpunct}{\mcitedefaultseppunct}\relax
\EndOfBibitem
\bibitem[Hanauer and K{\"o}hn(2011)Hanauer, and K{\"o}hn]{IC_MRCC_Kohn}
Hanauer,~M.; K{\"o}hn,~A. Pilot applications of internally contracted
  multireference coupled cluster theory, and how to choose the cluster operator
  properly. \emph{J. Chem. Phys.} \textbf{2011}, \emph{134}, 204111\relax
\mciteBstWouldAddEndPuncttrue
\mciteSetBstMidEndSepPunct{\mcitedefaultmidpunct}
{\mcitedefaultendpunct}{\mcitedefaultseppunct}\relax
\EndOfBibitem
\bibitem[Debashis~Mukherjee and Mukhopadhyay(1977)Debashis~Mukherjee, and
  Mukhopadhyay]{IC_MRCC_Mukherjee}
Debashis~Mukherjee,~R. K.~M.; Mukhopadhyay,~A. Applications of a
  non-perturbative many-body formalism to general open-shell atomic and
  molecular problems: calculation of the ground and the lowest $\pi-\pi^*$
  singlet and triplet energies and the first ionization potential of
  trans-butadiene. \emph{Mol. Phys.} \textbf{1977}, \emph{33}, 955--969\relax
\mciteBstWouldAddEndPuncttrue
\mciteSetBstMidEndSepPunct{\mcitedefaultmidpunct}
{\mcitedefaultendpunct}{\mcitedefaultseppunct}\relax
\EndOfBibitem
\bibitem[Yanai and Chan(2006)Yanai, and Chan]{IC_MRCC_Yanai}
Yanai,~T.; Chan,~G.~K. Canonical transformation theory for multireference
  problems. \emph{J. Chem. Phys.} \textbf{2006}, \emph{124}\relax
\mciteBstWouldAddEndPuncttrue
\mciteSetBstMidEndSepPunct{\mcitedefaultmidpunct}
{\mcitedefaultendpunct}{\mcitedefaultseppunct}\relax
\EndOfBibitem
\bibitem[Evangelista and Gauss(2011)Evangelista, and
  Gauss]{IC_MRCC_Evangelista}
Evangelista,~F.~A.; Gauss,~J. An orbital-invariant internally contracted
  multireference coupled cluster approach. \emph{J. Chem. Phys.} \textbf{2011},
  \emph{134}\relax
\mciteBstWouldAddEndPuncttrue
\mciteSetBstMidEndSepPunct{\mcitedefaultmidpunct}
{\mcitedefaultendpunct}{\mcitedefaultseppunct}\relax
\EndOfBibitem
\bibitem[Barkoutsos \latin{et~al.}(2018)Barkoutsos, Gonthier, Sokolov, Moll,
  Salis, Fuhrer, Ganzhorn, Egger, Troyer, Mezzacapo, Filipp, and
  Tavernelli]{Trotter_1}
Barkoutsos,~P.~K.; Gonthier,~J.~F.; Sokolov,~I.; Moll,~N.; Salis,~G.;
  Fuhrer,~A.; Ganzhorn,~M.; Egger,~D.~J.; Troyer,~M.; Mezzacapo,~A.;
  Filipp,~S.; Tavernelli,~I. Quantum algorithms for electronic structure
  calculations: Particle-hole Hamiltonian and optimized wave-function
  expansions. \emph{Phys. Rev. A} \textbf{2018}, \emph{98}, 022322\relax
\mciteBstWouldAddEndPuncttrue
\mciteSetBstMidEndSepPunct{\mcitedefaultmidpunct}
{\mcitedefaultendpunct}{\mcitedefaultseppunct}\relax
\EndOfBibitem
\bibitem[Moll \latin{et~al.}(2018)Moll, Barkoutsos, Bishop, Chow, Cross, Egger,
  Filipp, Fuhrer, Gambetta, Ganzhorn, Kandala, Mezzacapo, M{\"u}ller, Riess,
  Salis, Smolin, Tavernelli, and Temme]{Trotter_2}
Moll,~N.; Barkoutsos,~P.; Bishop,~L.~S.; Chow,~J.~M.; Cross,~A.; Egger,~D.~J.;
  Filipp,~S.; Fuhrer,~A.; Gambetta,~J.~M.; Ganzhorn,~M.; Kandala,~A.;
  Mezzacapo,~A.; M{\"u}ller,~P.; Riess,~W.; Salis,~G.; Smolin,~J.;
  Tavernelli,~I.; Temme,~K. Quantum optimization using variational algorithms
  on near-term quantum devices. \emph{Quantum Sci. Technol.} \textbf{2018},
  \emph{3}, 030503\relax
\mciteBstWouldAddEndPuncttrue
\mciteSetBstMidEndSepPunct{\mcitedefaultmidpunct}
{\mcitedefaultendpunct}{\mcitedefaultseppunct}\relax
\EndOfBibitem
\bibitem[Evangelista \latin{et~al.}(2019)Evangelista, Chan, and
  Scuseria]{Trotter_4}
Evangelista,~F.~A.; Chan,~G. K.-L.; Scuseria,~G.~E. Exact parameterization of
  fermionic wave functions via unitary coupled cluster theory. \emph{J. Chem.
  Phys.} \textbf{2019}, \emph{151}, 244112\relax
\mciteBstWouldAddEndPuncttrue
\mciteSetBstMidEndSepPunct{\mcitedefaultmidpunct}
{\mcitedefaultendpunct}{\mcitedefaultseppunct}\relax
\EndOfBibitem
\bibitem[Sokolov \latin{et~al.}(2020)Sokolov, Barkoutsos, Ollitrault,
  Greenberg, Rice, Pistoia, and Tavernelli]{Trotter_5}
Sokolov,~I.~O.; Barkoutsos,~P.~K.; Ollitrault,~P.~J.; Greenberg,~D.; Rice,~J.;
  Pistoia,~M.; Tavernelli,~I. Quantum orbital-optimized unitary coupled cluster
  methods in the strongly correlated regime: Can quantum algorithms outperform
  their classical equivalents? \emph{J. Chem. Phys.} \textbf{2020}, \emph{152},
  124107\relax
\mciteBstWouldAddEndPuncttrue
\mciteSetBstMidEndSepPunct{\mcitedefaultmidpunct}
{\mcitedefaultendpunct}{\mcitedefaultseppunct}\relax
\EndOfBibitem
\bibitem[Sugisaki \latin{et~al.}(2022)Sugisaki, Kato, Minato, Okuwaki, and
  Mochizuki]{MRUCCSD}
Sugisaki,~K.; Kato,~T.; Minato,~Y.; Okuwaki,~K.; Mochizuki,~Y. Variational
  quantum eigensolver simulations with the multireference unitary coupled
  cluster ansatz: a case study of the C2v quasi-reaction pathway of beryllium
  insertion into a H2 molecule. \emph{Phys. Chem. Chem. Phys.} \textbf{2022},
  \emph{24}, 8439--8452\relax
\mciteBstWouldAddEndPuncttrue
\mciteSetBstMidEndSepPunct{\mcitedefaultmidpunct}
{\mcitedefaultendpunct}{\mcitedefaultseppunct}\relax
\EndOfBibitem
\bibitem[McClean \latin{et~al.}(2017)McClean, Kimchi-Schwartz, Carter, and
  de~Jong]{McClean17}
McClean,~J.~R.; Kimchi-Schwartz,~M.~E.; Carter,~J.; de~Jong,~W.~A. Hybrid
  quantum-classical hierarchy for mitigation of decoherence and determination
  of excited states. \emph{Phys. Rev. A} \textbf{2017}, \emph{95}, 042308\relax
\mciteBstWouldAddEndPuncttrue
\mciteSetBstMidEndSepPunct{\mcitedefaultmidpunct}
{\mcitedefaultendpunct}{\mcitedefaultseppunct}\relax
\EndOfBibitem
\bibitem[Colless \latin{et~al.}(2018)Colless, Ramasesh, Dahlen, Blok,
  Kimchi-Schwartz, McClean, Carter, de~Jong, and Siddiqi]{Colless18}
Colless,~J.~I.; Ramasesh,~V.~V.; Dahlen,~D.; Blok,~M.~S.;
  Kimchi-Schwartz,~M.~E.; McClean,~J.~R.; Carter,~J.; de~Jong,~W.~A.;
  Siddiqi,~I. Computation of Molecular Spectra on a Quantum Processor with an
  Error-Resilient Algorithm. \emph{Phys. Rev. X} \textbf{2018}, \emph{8},
  011021\relax
\mciteBstWouldAddEndPuncttrue
\mciteSetBstMidEndSepPunct{\mcitedefaultmidpunct}
{\mcitedefaultendpunct}{\mcitedefaultseppunct}\relax
\EndOfBibitem
\bibitem[Higgott \latin{et~al.}(2019)Higgott, Wang, and Brierley]{Higgott19}
Higgott,~O.; Wang,~D.; Brierley,~S. Variational {Q}uantum {C}omputation of
  {E}xcited {S}tates. \emph{Quantum} \textbf{2019}, \emph{3}, 156\relax
\mciteBstWouldAddEndPuncttrue
\mciteSetBstMidEndSepPunct{\mcitedefaultmidpunct}
{\mcitedefaultendpunct}{\mcitedefaultseppunct}\relax
\EndOfBibitem
\bibitem[Nakanishi \latin{et~al.}(2019)Nakanishi, Mitarai, and
  Fujii]{Nakanishi19}
Nakanishi,~K.~M.; Mitarai,~K.; Fujii,~K. Subspace-search variational quantum
  eigensolver for excited states. \emph{Phys. Rev. Res.} \textbf{2019},
  \emph{1}, 033062\relax
\mciteBstWouldAddEndPuncttrue
\mciteSetBstMidEndSepPunct{\mcitedefaultmidpunct}
{\mcitedefaultendpunct}{\mcitedefaultseppunct}\relax
\EndOfBibitem
\bibitem[Ollitrault \latin{et~al.}(2020)Ollitrault, Kandala, Chen, Barkoutsos,
  Mez~zacapo, Pistoia, Sheldon, Woerner, Gambetta, and
  Tavernelli]{Ollitrault20}
Ollitrault,~P.~J.; Kandala,~A.; Chen,~C.-F.; Barkoutsos,~P.~K.; Mez~zacapo,~A.;
  Pistoia,~M.; Sheldon,~S.; Woerner,~S.; Gambetta,~J.~M.; Tavernelli,~I.~v.
  Quantum equation of motion for computing molecular excitation energies on a
  noisy quantum processor. \emph{Phys. Rev. Res.} \textbf{2020}, \emph{2},
  043140\relax
\mciteBstWouldAddEndPuncttrue
\mciteSetBstMidEndSepPunct{\mcitedefaultmidpunct}
{\mcitedefaultendpunct}{\mcitedefaultseppunct}\relax
\EndOfBibitem
\bibitem[Zhang \latin{et~al.}(2021)Zhang, Gomes, Yao, Orth, and
  Iadecola]{Zhang21A}
Zhang,~F.; Gomes,~N.; Yao,~Y.; Orth,~P.~P.; Iadecola,~T. Adaptive variational
  quantum eigensolvers for highly excited states. \emph{Phys. Rev. B}
  \textbf{2021}, \emph{104}, 075159\relax
\mciteBstWouldAddEndPuncttrue
\mciteSetBstMidEndSepPunct{\mcitedefaultmidpunct}
{\mcitedefaultendpunct}{\mcitedefaultseppunct}\relax
\EndOfBibitem
\bibitem[Tkachenko \latin{et~al.}(2022)Tkachenko, Zhang, Cincio, Boldyrev,
  Tretiak, and Dub]{Tkachenko22}
Tkachenko,~N.~V.; Zhang,~Y.; Cincio,~L.; Boldyrev,~A.~I.; Tretiak,~S.;
  Dub,~P.~A. Quantum Davidson Algorithm for Excited States. \textbf{2022},
  arXiv:2204.10741 [quant--ph], date of access: 2023/9/29\relax
\mciteBstWouldAddEndPuncttrue
\mciteSetBstMidEndSepPunct{\mcitedefaultmidpunct}
{\mcitedefaultendpunct}{\mcitedefaultseppunct}\relax
\EndOfBibitem
\bibitem[Heya \latin{et~al.}(2023)Heya, Nakanishi, Mitarai, Yan, Zuo, Suzuki,
  Sugiyama, Tamate, Tabuchi, Fujii, and Nakamura]{Heya23}
Heya,~K.; Nakanishi,~K.~M.; Mitarai,~K.; Yan,~Z.; Zuo,~K.; Suzuki,~Y.;
  Sugiyama,~T.; Tamate,~S.; Tabuchi,~Y.; Fujii,~K.; Nakamura,~Y. Subspace
  variational quantum simulator. \emph{Phys. Rev. Res.} \textbf{2023},
  \emph{5}, 023078\relax
\mciteBstWouldAddEndPuncttrue
\mciteSetBstMidEndSepPunct{\mcitedefaultmidpunct}
{\mcitedefaultendpunct}{\mcitedefaultseppunct}\relax
\EndOfBibitem
\bibitem[Tsuchimochi \latin{et~al.}(2023)Tsuchimochi, Ryo, Ten-no, and
  Sasasako]{Tsuchimochi23A}
Tsuchimochi,~T.; Ryo,~Y.; Ten-no,~S.~L.; Sasasako,~K. Improved Algorithms of
  Quantum Imaginary Time Evolution for Ground and Excited States of Molecular
  Syst ems. \emph{J. Chem. Theory Comput.} \textbf{2023}, \emph{19},
  503--513\relax
\mciteBstWouldAddEndPuncttrue
\mciteSetBstMidEndSepPunct{\mcitedefaultmidpunct}
{\mcitedefaultendpunct}{\mcitedefaultseppunct}\relax
\EndOfBibitem
\bibitem[Tsuchimochi \latin{et~al.}(2022)Tsuchimochi, Ryo, and
  Ten-no]{Tsuchimochi23B}
Tsuchimochi,~T.; Ryo,~Y.; Ten-no,~S.~L. Multi-state quantum simulations via
  model-space quantum imaginary time evolution. \textbf{2022}, arXiv:2206.04494
  [quant--ph], date of access: 2023/9/29\relax
\mciteBstWouldAddEndPuncttrue
\mciteSetBstMidEndSepPunct{\mcitedefaultmidpunct}
{\mcitedefaultendpunct}{\mcitedefaultseppunct}\relax
\EndOfBibitem
\bibitem[Takahiro \latin{et~al.}(2023)Takahiro, Ten-no, and
  Tsuchimochi]{Yoshikura23}
Takahiro,~Y.; Ten-no,~S.~L.; Tsuchimochi,~T. Quantum Inverse Algorithm via
  Adaptive Variational Quantum Linear Solver: Applications to General
  Eigenstates. \emph{J. Phys. Chem. A} \textbf{2023}, \emph{127},
  6577--6592\relax
\mciteBstWouldAddEndPuncttrue
\mciteSetBstMidEndSepPunct{\mcitedefaultmidpunct}
{\mcitedefaultendpunct}{\mcitedefaultseppunct}\relax
\EndOfBibitem
\bibitem[Sokolov \latin{et~al.}(2020)Sokolov, Barkoutsos, Ollitrault,
  Greenberg, Rice, Pistoia, and Tavernelli]{Sokolov20}
Sokolov,~I.~O.; Barkoutsos,~P.~K.; Ollitrault,~P.~J.; Greenberg,~D.; Rice,~J.;
  Pistoia,~M.; Tavernelli,~I. Quantum orbital-optimized unitary coupled cluster
  methods in the strongly correlated regime: Can quantum algorithms outperform
  their classical equivalents? \emph{J. Chem. Phys.} \textbf{2020}, \emph{152},
  124107\relax
\mciteBstWouldAddEndPuncttrue
\mciteSetBstMidEndSepPunct{\mcitedefaultmidpunct}
{\mcitedefaultendpunct}{\mcitedefaultseppunct}\relax
\EndOfBibitem
\bibitem[Mizukami \latin{et~al.}(2020)Mizukami, Mitarai, Nakagawa, Yamamoto,
  Yan, and Ohnishi]{Mizukami20}
Mizukami,~W.; Mitarai,~K.; Nakagawa,~Y.~O.; Yamamoto,~T.; Yan,~T.;
  Ohnishi,~Y.-y. Orbital optimized unitary coupled cluster theory for quantum
  computer. \emph{Phys. Rev. Research} \textbf{2020}, \emph{2}, 033421\relax
\mciteBstWouldAddEndPuncttrue
\mciteSetBstMidEndSepPunct{\mcitedefaultmidpunct}
{\mcitedefaultendpunct}{\mcitedefaultseppunct}\relax
\EndOfBibitem
\bibitem[Tsuchimochi \latin{et~al.}(2022)Tsuchimochi, Mori, Shimomoto,
  Nishimaki, Ryo, Taii, Yoshikura, Chung, Sasasako, and Yoshimura]{quket}
Tsuchimochi,~T.; Mori,~Y.; Shimomoto,~Y.; Nishimaki,~T.; Ryo,~Y.; Taii,~M.;
  Yoshikura,~T.; Chung,~T.~S.; Sasasako,~K.; Yoshimura,~K. Quket: the
  comprehensive quantum simulator for quantum chemistry. 2022;
  \url{https://github.com/quket/quket}, date of access: 2023/9/29\relax
\mciteBstWouldAddEndPuncttrue
\mciteSetBstMidEndSepPunct{\mcitedefaultmidpunct}
{\mcitedefaultendpunct}{\mcitedefaultseppunct}\relax
\EndOfBibitem
\bibitem[McClean \latin{et~al.}(2020)McClean, Rubin, Sung, Kivlichan,
  Bonet-Monroig, Cao, Dai, Fried, Gidney, Gimby, Gokhale, H{\"a}ner, Hardikar,
  Havl{\'i}{\v{c}}ek, Higgott, Huang, Izaac, Jiang, Liu, McArdle, Neeley,
  O'Brien, O'Gorman, Ozfidan, Radin, Romero, Sawaya, Senjean, Setia, Sim,
  Steiger, Steudtner, Sun, Sun, Wang, Zhang, and Babbush]{OPENFERMION}
McClean,~J.~R.; Rubin,~N.~C.; Sung,~K.~J.; Kivlichan,~I.~D.; Bonet-Monroig,~X.;
  Cao,~Y.; Dai,~C.; Fried,~E.~S.; Gidney,~C.; Gimby,~B.; Gokhale,~P.;
  H{\"a}ner,~T.; Hardikar,~T.; Havl{\'i}{\v{c}}ek,~V.; Higgott,~O.; Huang,~C.;
  Izaac,~J.; Jiang,~Z.; Liu,~X.; McArdle,~S.; Neeley,~M.; O'Brien,~T.;
  O'Gorman,~B.; Ozfidan,~I.; Radin,~M.~D.; Romero,~J.; Sawaya,~N. P.~D.;
  Senjean,~B.; Setia,~K.; Sim,~S.; Steiger,~D.~S.; Steudtner,~M.; Sun,~Q.;
  Sun,~W.; Wang,~D.; Zhang,~F.; Babbush,~R. OpenFermion: the electronic
  structure package for quantum computers. \emph{Quantum Sci. Technol.}
  \textbf{2020}, \emph{5}, 034014\relax
\mciteBstWouldAddEndPuncttrue
\mciteSetBstMidEndSepPunct{\mcitedefaultmidpunct}
{\mcitedefaultendpunct}{\mcitedefaultseppunct}\relax
\EndOfBibitem
\bibitem[Sun \latin{et~al.}(2018)Sun, Berkelbach, Blunt, Booth, Guo, Li, Liu,
  McClain, Sayfutyarova, Sharma, Wouters, and Chan]{PYSCF}
Sun,~Q.; Berkelbach,~T.~C.; Blunt,~N.~S.; Booth,~G.~H.; Guo,~S.; Li,~Z.;
  Liu,~J.; McClain,~J.~D.; Sayfutyarova,~E.~R.; Sharma,~S.; Wouters,~S.;
  Chan,~G. K.-L. PySCF: the Python-based simulations of chemistry framework.
  \emph{WIREs Comput Mol Sci.} \textbf{2018}, \emph{8}, e1340\relax
\mciteBstWouldAddEndPuncttrue
\mciteSetBstMidEndSepPunct{\mcitedefaultmidpunct}
{\mcitedefaultendpunct}{\mcitedefaultseppunct}\relax
\EndOfBibitem
\bibitem[Suzuki \latin{et~al.}(2021)Suzuki, Kawase, Masumura, Hiraga, Nakadai,
  Chen, Nakanishi, Mitarai, Imai, Tamiya, Yamamoto, Yan, Kawakubo, Nakagawa,
  Ibe, Zhang, Yamashita, Yoshimura, Hayashi, and Fujii]{qulacs}
Suzuki,~Y.; Kawase,~Y.; Masumura,~Y.; Hiraga,~Y.; Nakadai,~M.; Chen,~J.;
  Nakanishi,~K.~M.; Mitarai,~K.; Imai,~R.; Tamiya,~S.; Yamamoto,~T.; Yan,~T.;
  Kawakubo,~T.; Nakagawa,~Y.~O.; Ibe,~Y.; Zhang,~Y.; Yamashita,~H.;
  Yoshimura,~H.; Hayashi,~A.; Fujii,~K. Qulacs: a fast and versatile quantum
  circuit simulator for research purpose. \emph{Quantum} \textbf{2021},
  \emph{5}, 559\relax
\mciteBstWouldAddEndPuncttrue
\mciteSetBstMidEndSepPunct{\mcitedefaultmidpunct}
{\mcitedefaultendpunct}{\mcitedefaultseppunct}\relax
\EndOfBibitem
\bibitem[Jordan and Wigner(1928)Jordan, and Wigner]{JW_trans}
Jordan,~P.; Wigner,~E. \emph{Z. Phys.} \textbf{1928}, \emph{47}, 631\relax
\mciteBstWouldAddEndPuncttrue
\mciteSetBstMidEndSepPunct{\mcitedefaultmidpunct}
{\mcitedefaultendpunct}{\mcitedefaultseppunct}\relax
\EndOfBibitem
\bibitem[Bravyi \latin{et~al.}(2017)Bravyi, Gambetta, Mezzacapo, and
  Temme]{bravyi17}
Bravyi,~S.; Gambetta,~J.~M.; Mezzacapo,~A.; Temme,~K. Tapering off qubits to
  simulate fermionic Hamiltonians. \textbf{2017}, arXiv:1701.08213 [quant--ph],
  date of access: 2023/9/29\relax
\mciteBstWouldAddEndPuncttrue
\mciteSetBstMidEndSepPunct{\mcitedefaultmidpunct}
{\mcitedefaultendpunct}{\mcitedefaultseppunct}\relax
\EndOfBibitem
\bibitem[Setia \latin{et~al.}(2020)Setia, Chen, Rice, Mezzacapo, Pistoia, and
  Whitfield]{Setia20}
Setia,~K.; Chen,~R.; Rice,~J.~E.; Mezzacapo,~A.; Pistoia,~M.; Whitfield,~J.~D.
  Reducing Qubit Requirements for Quantum Simulations Using Molecular Point
  Group Symmetries. \emph{J. Chem. Theory Comput.} \textbf{2020}, \emph{16},
  6091--6097\relax
\mciteBstWouldAddEndPuncttrue
\mciteSetBstMidEndSepPunct{\mcitedefaultmidpunct}
{\mcitedefaultendpunct}{\mcitedefaultseppunct}\relax
\EndOfBibitem
\bibitem[Werner and Knowles(1988)Werner, and Knowles]{MRCI_1}
Werner,~H.; Knowles,~P.~J. {An efficient internally contracted
  multiconfiguration–reference configuration interaction method}. \emph{J.
  Chem. Phys.} \textbf{1988}, \emph{89}, 5803--5814\relax
\mciteBstWouldAddEndPuncttrue
\mciteSetBstMidEndSepPunct{\mcitedefaultmidpunct}
{\mcitedefaultendpunct}{\mcitedefaultseppunct}\relax
\EndOfBibitem
\bibitem[MRC(1988)]{MRCI_2}
An efficient method for the evaluation of coupling coefficients in
  configuration interaction calculations. \emph{Chem. Phys. Lett.}
  \textbf{1988}, \emph{145}, 514--522\relax
\mciteBstWouldAddEndPuncttrue
\mciteSetBstMidEndSepPunct{\mcitedefaultmidpunct}
{\mcitedefaultendpunct}{\mcitedefaultseppunct}\relax
\EndOfBibitem
\bibitem[Pople \latin{et~al.}(1977)Pople, Seeger, and Krishnan]{Pople}
Pople,~J.~A.; Seeger,~R.; Krishnan,~R. Variational configuration interaction
  methods and comparison with perturbation theory. \emph{Int. J. Quantum Chem.}
  \textbf{1977}, \emph{12}, 149--163\relax
\mciteBstWouldAddEndPuncttrue
\mciteSetBstMidEndSepPunct{\mcitedefaultmidpunct}
{\mcitedefaultendpunct}{\mcitedefaultseppunct}\relax
\EndOfBibitem
\bibitem[Li and Paldus(2003)Li, and Paldus]{LiH}
Li,~X.; Paldus,~J. The general-model-space state-universal coupled-cluster
  method exemplified by the LiH molecule. \emph{J. Chem. Phys.} \textbf{2003},
  \emph{119}, 5346--5357\relax
\mciteBstWouldAddEndPuncttrue
\mciteSetBstMidEndSepPunct{\mcitedefaultmidpunct}
{\mcitedefaultendpunct}{\mcitedefaultseppunct}\relax
\EndOfBibitem
\bibitem[Werner \latin{et~al.}(2020)Werner, Knowles, Manby, Black, Doll,
  Heßelmann, Kats, Köhn, Korona, Kreplin, Ma, Miller, Mitrushchenkov,
  Peterson, Polyak, Rauhut, and Sibaev]{molpro}
Werner,~H.-J.; Knowles,~P.~J.; Manby,~F.~R.; Black,~J.~A.; Doll,~K.;
  Heßelmann,~A.; Kats,~D.; Köhn,~A.; Korona,~T.; Kreplin,~D.~A.; Ma,~Q.;
  Miller,~I.,~Thomas~F.; Mitrushchenkov,~A.; Peterson,~K.~A.; Polyak,~I.;
  Rauhut,~G.; Sibaev,~M. {The Molpro quantum chemistry package}. \emph{J. Chem.
  Phys.} \textbf{2020}, \emph{152}\relax
\mciteBstWouldAddEndPuncttrue
\mciteSetBstMidEndSepPunct{\mcitedefaultmidpunct}
{\mcitedefaultendpunct}{\mcitedefaultseppunct}\relax
\EndOfBibitem
\bibitem[Loos \latin{et~al.}(2018)Loos, Scemama, Blondel, Garniron, Caffarel,
  and Jacquemin]{H2O}
Loos,~P.-F.; Scemama,~A.; Blondel,~A.; Garniron,~Y.; Caffarel,~M.;
  Jacquemin,~D. A Mountaineering Strategy to Excited States: Highly Accurate
  Reference Energies and Benchmarks. \emph{J. Chem. Theory Comput.}
  \textbf{2018}, \emph{14}, 4360--4379\relax
\mciteBstWouldAddEndPuncttrue
\mciteSetBstMidEndSepPunct{\mcitedefaultmidpunct}
{\mcitedefaultendpunct}{\mcitedefaultseppunct}\relax
\EndOfBibitem
\end{mcitethebibliography}

\providecommand{\noopsort}[1]{}\providecommand{\singleletter}[1]{#1}%
\providecommand{\latin}[1]{#1}
\makeatletter
\providecommand{\doi}
  {\begingroup\let\do\@makeother\dospecials
  \catcode`\{=1 \catcode`\}=2 \doi@aux}
\providecommand{\doi@aux}[1]{\endgroup\texttt{#1}}
\makeatother
\providecommand*\mcitethebibliography{\thebibliography}
\csname @ifundefined\endcsname{endmcitethebibliography}
  {\let\endmcitethebibliography\endthebibliography}{}

\end{document}